\documentclass[letterpaper,compsoc,twoside,english]{IEEEtran}
\usepackage{fixltx2e} 
\usepackage{cmap} 
\usepackage[T1]{fontenc}
\usepackage[utf8]{inputenc}
\usepackage{ifthen}
\usepackage{babel}
\usepackage{amsmath}

\usepackage[font={small,it},labelfont=bf]{caption}
\usepackage{float}

\usepackage{graphicx}

\usepackage{mathptmx}
\usepackage[scaled=.90]{helvet}
\usepackage{courier}

\makeatletter
\g@addto@macro\@verbatim\footnotesize
\makeatother

\usepackage{amsmath}
\usepackage{amsfonts}
\usepackage{bm}

\usepackage{color}

\definecolor{orange}{cmyk}{0,0.4,0.8,0.2}
\definecolor{darkorange}{rgb}{.71,0.21,0.01}
\definecolor{darkblue}{rgb}{.01,0.21,0.71}
\definecolor{darkgreen}{rgb}{.1,.52,.09}

\usepackage{hyperref}
\hypersetup{pdftex,  
  breaklinks=true,  
  colorlinks=true,
  urlcolor=blue,
  linkcolor=darkblue,
  citecolor=darkgreen,
  }

\usepackage{graphicx}

\ifthenelse{\isundefined{\longtable}}{}{
  \renewenvironment{longtable}{\begin{center}\begin{tabular}}%
    {\end{tabular}\end{center}\vspace{2mm}}
}

\usepackage{fancyvrb}

\makeatletter
\def\PY@reset{\let\PY@it=\relax \let\PY@bf=\relax%
    \let\PY@ul=\relax \let\PY@tc=\relax%
    \let\PY@bc=\relax \let\PY@ff=\relax}
\def\PY@tok#1{\csname PY@tok@#1\endcsname}
\def\PY@toks#1+{\ifx\relax#1\empty\else%
    \PY@tok{#1}\expandafter\PY@toks\fi}
\def\PY@do#1{\PY@bc{\PY@tc{\PY@ul{%
    \PY@it{\PY@bf{\PY@ff{#1}}}}}}}
\def\PY#1#2{\PY@reset\PY@toks#1+\relax+\PY@do{#2}}

\def\PY@tok@gu{\let\PY@bf=\textbf\def\PY@tc##1{\textcolor[rgb]{0.50,0.00,0.50}{##1}}}
\def\PY@tok@gt{\def\PY@tc##1{\textcolor[rgb]{0.00,0.25,0.82}{##1}}}
\def\PY@tok@gs{\let\PY@bf=\textbf}
\def\PY@tok@gr{\def\PY@tc##1{\textcolor[rgb]{1.00,0.00,0.00}{##1}}}
\def\PY@tok@cm{\let\PY@it=\textit\def\PY@tc##1{\textcolor[rgb]{0.25,0.50,0.56}{##1}}}
\def\PY@tok@vg{\def\PY@tc##1{\textcolor[rgb]{0.73,0.38,0.84}{##1}}}
\def\PY@tok@m{\def\PY@tc##1{\textcolor[rgb]{0.13,0.50,0.31}{##1}}}
\def\PY@tok@mh{\def\PY@tc##1{\textcolor[rgb]{0.13,0.50,0.31}{##1}}}
\def\PY@tok@go{\def\PY@tc##1{\textcolor[rgb]{0.19,0.19,0.19}{##1}}}
\def\PY@tok@ge{\let\PY@it=\textit}
\def\PY@tok@gd{\def\PY@tc##1{\textcolor[rgb]{0.63,0.00,0.00}{##1}}}
\def\PY@tok@il{\def\PY@tc##1{\textcolor[rgb]{0.13,0.50,0.31}{##1}}}
\def\PY@tok@cs{\def\PY@tc##1{\textcolor[rgb]{0.25,0.50,0.56}{##1}}\def\PY@bc##1{\colorbox[rgb]{1.00,0.94,0.94}{##1}}}
\def\PY@tok@cp{\def\PY@tc##1{\textcolor[rgb]{0.00,0.44,0.13}{##1}}}
\def\PY@tok@gi{\def\PY@tc##1{\textcolor[rgb]{0.00,0.63,0.00}{##1}}}
\def\PY@tok@gh{\let\PY@bf=\textbf\def\PY@tc##1{\textcolor[rgb]{0.00,0.00,0.50}{##1}}}
\def\PY@tok@ni{\let\PY@bf=\textbf\def\PY@tc##1{\textcolor[rgb]{0.84,0.33,0.22}{##1}}}
\def\PY@tok@nl{\let\PY@bf=\textbf\def\PY@tc##1{\textcolor[rgb]{0.00,0.13,0.44}{##1}}}
\def\PY@tok@nn{\let\PY@bf=\textbf\def\PY@tc##1{\textcolor[rgb]{0.05,0.52,0.71}{##1}}}
\def\PY@tok@no{\def\PY@tc##1{\textcolor[rgb]{0.38,0.68,0.84}{##1}}}
\def\PY@tok@na{\def\PY@tc##1{\textcolor[rgb]{0.25,0.44,0.63}{##1}}}
\def\PY@tok@nb{\def\PY@tc##1{\textcolor[rgb]{0.00,0.44,0.13}{##1}}}
\def\PY@tok@nc{\let\PY@bf=\textbf\def\PY@tc##1{\textcolor[rgb]{0.05,0.52,0.71}{##1}}}
\def\PY@tok@nd{\let\PY@bf=\textbf\def\PY@tc##1{\textcolor[rgb]{0.33,0.33,0.33}{##1}}}
\def\PY@tok@ne{\def\PY@tc##1{\textcolor[rgb]{0.00,0.44,0.13}{##1}}}
\def\PY@tok@nf{\def\PY@tc##1{\textcolor[rgb]{0.02,0.16,0.49}{##1}}}
\def\PY@tok@si{\let\PY@it=\textit\def\PY@tc##1{\textcolor[rgb]{0.44,0.63,0.82}{##1}}}
\def\PY@tok@s2{\def\PY@tc##1{\textcolor[rgb]{0.25,0.44,0.63}{##1}}}
\def\PY@tok@vi{\def\PY@tc##1{\textcolor[rgb]{0.73,0.38,0.84}{##1}}}
\def\PY@tok@nt{\let\PY@bf=\textbf\def\PY@tc##1{\textcolor[rgb]{0.02,0.16,0.45}{##1}}}
\def\PY@tok@nv{\def\PY@tc##1{\textcolor[rgb]{0.73,0.38,0.84}{##1}}}
\def\PY@tok@s1{\def\PY@tc##1{\textcolor[rgb]{0.25,0.44,0.63}{##1}}}
\def\PY@tok@vc{\def\PY@tc##1{\textcolor[rgb]{0.73,0.38,0.84}{##1}}}
\def\PY@tok@sh{\def\PY@tc##1{\textcolor[rgb]{0.25,0.44,0.63}{##1}}}
\def\PY@tok@ow{\let\PY@bf=\textbf\def\PY@tc##1{\textcolor[rgb]{0.00,0.44,0.13}{##1}}}
\def\PY@tok@mf{\def\PY@tc##1{\textcolor[rgb]{0.13,0.50,0.31}{##1}}}
\def\PY@tok@bp{\def\PY@tc##1{\textcolor[rgb]{0.00,0.44,0.13}{##1}}}
\def\PY@tok@c1{\let\PY@it=\textit\def\PY@tc##1{\textcolor[rgb]{0.25,0.50,0.56}{##1}}}
\def\PY@tok@kc{\let\PY@bf=\textbf\def\PY@tc##1{\textcolor[rgb]{0.00,0.44,0.13}{##1}}}
\def\PY@tok@c{\let\PY@it=\textit\def\PY@tc##1{\textcolor[rgb]{0.25,0.50,0.56}{##1}}}
\def\PY@tok@sx{\def\PY@tc##1{\textcolor[rgb]{0.78,0.36,0.04}{##1}}}
\def\PY@tok@err{\def\PY@bc##1{\fcolorbox[rgb]{1.00,0.00,0.00}{1,1,1}{##1}}}
\def\PY@tok@kd{\let\PY@bf=\textbf\def\PY@tc##1{\textcolor[rgb]{0.00,0.44,0.13}{##1}}}
\def\PY@tok@ss{\def\PY@tc##1{\textcolor[rgb]{0.32,0.47,0.09}{##1}}}
\def\PY@tok@sr{\def\PY@tc##1{\textcolor[rgb]{0.14,0.33,0.53}{##1}}}
\def\PY@tok@mo{\def\PY@tc##1{\textcolor[rgb]{0.13,0.50,0.31}{##1}}}
\def\PY@tok@kn{\let\PY@bf=\textbf\def\PY@tc##1{\textcolor[rgb]{0.00,0.44,0.13}{##1}}}
\def\PY@tok@mi{\def\PY@tc##1{\textcolor[rgb]{0.13,0.50,0.31}{##1}}}
\def\PY@tok@gp{\let\PY@bf=\textbf\def\PY@tc##1{\textcolor[rgb]{0.78,0.36,0.04}{##1}}}
\def\PY@tok@o{\def\PY@tc##1{\textcolor[rgb]{0.40,0.40,0.40}{##1}}}
\def\PY@tok@kr{\let\PY@bf=\textbf\def\PY@tc##1{\textcolor[rgb]{0.00,0.44,0.13}{##1}}}
\def\PY@tok@s{\def\PY@tc##1{\textcolor[rgb]{0.25,0.44,0.63}{##1}}}
\def\PY@tok@kp{\def\PY@tc##1{\textcolor[rgb]{0.00,0.44,0.13}{##1}}}
\def\PY@tok@w{\def\PY@tc##1{\textcolor[rgb]{0.73,0.73,0.73}{##1}}}
\def\PY@tok@kt{\def\PY@tc##1{\textcolor[rgb]{0.56,0.13,0.00}{##1}}}
\def\PY@tok@sc{\def\PY@tc##1{\textcolor[rgb]{0.25,0.44,0.63}{##1}}}
\def\PY@tok@sb{\def\PY@tc##1{\textcolor[rgb]{0.25,0.44,0.63}{##1}}}
\def\PY@tok@k{\let\PY@bf=\textbf\def\PY@tc##1{\textcolor[rgb]{0.00,0.44,0.13}{##1}}}
\def\PY@tok@se{\let\PY@bf=\textbf\def\PY@tc##1{\textcolor[rgb]{0.25,0.44,0.63}{##1}}}
\def\PY@tok@sd{\let\PY@it=\textit\def\PY@tc##1{\textcolor[rgb]{0.25,0.44,0.63}{##1}}}

\def\PYZbs{\char`\\}
\def\PYZus{\char`\_}

\def\PYZsh{\char`\#}
\def\PYZpc{\char`\%}

\makeatother

\providecommand*{\DUrole}[2]{%
  \ifcsname DUrole#1\endcsname%
    \csname DUrole#1\endcsname{#2}%
  \else
    \ifcsname docutilsrole#1\endcsname%
      \csname docutilsrole#1\endcsname{#2}%
    \else%
      #2%
    \fi%
  \fi%
}

\ifthenelse{\isundefined{\hypersetup}}{
  \usepackage[unicode,colorlinks=true,linkcolor=blue,urlcolor=blue]{hyperref}
  \urlstyle{same} 
}{}

\begin{document}
\title{Building a Framework for Predictive Science}\author{Michael M. McKerns, Leif Strand, Tim Sullivan, Alta Fang, Michael A.G. Aivazis\thanks{
The corresponding author is with California Institute of Technology, e-mail: \protect\href{mailto:mmckerns@caltech.edu}{mmckerns@caltech.edu}.
        }}\maketitle
        \renewcommand{\leftmark}{PROC. OF THE 10th PYTHON IN SCIENCE CONF. (SCIPY 2011)}
        \renewcommand{\rightmark}{BUILDING A FRAMEWORK FOR PREDICTIVE SCIENCE}

\newcommand*{\docutilsroleref}{\ref}
\newcommand*{\docutilsrolelabel}{\label}

\begin{abstract}Key questions that scientists and engineers typically want to address can be formulated in terms of predictive science. Questions such as: ``How well does my computational model represent reality?'', ``What are the most important parameters in the problem?'', and ``What is the best next experiment to perform?'' are fundamental in solving scientific problems. \texttt{mystic} is a framework for massively-parallel optimization and rigorous sensitivity analysis that enables these motivating questions to be addressed quantitatively as global optimization problems. Often realistic physics, engineering, and materials models may have hundreds of input parameters, hundreds of constraints, and may require execution times of seconds or longer. In more extreme cases, realistic models may be multi-scale, and require the use of high-performance computing clusters for their evaluation. Predictive calculations, formulated as a global optimization over a potential surface in design parameter space, may require an already prohibitively large simulation to be performed hundreds, if not thousands, of times. The need to prepare, schedule, and monitor thousands of model evaluations, and dynamically explore and analyze results, is a challenging problem that requires a software infrastructure capable of distributing and managing computations on large-scale heterogeneous resources.  In this paper, we present the design behind an optimization framework, and also a framework for heterogeneous computing, that when utilized together, can make computationally intractable sensitivity and optimization problems much more tractable. The optimization framework provides global search algorithms that have been extended to parallel, where evaluations of the model can be distributed to appropriate large-scale resources, while the optimizer centrally manages their interactions and navigates the objective function.  New methods have been developed for imposing and solving constraints that aid in reducing the size and complexity of the optimization problem. Additionally, new algorithms have been developed that launch multiple optimizers in parallel, thus allowing highly efficient local search algorithms to provide fast global optimization. In this way, parallelism in optimization also can allow us to not only find global minima, but to simultaneously find all local minima and transition points -{}- thus providing a much more efficient means of mapping out a potential energy surface.\end{abstract}


\subsection*{Introduction%
  \phantomsection%
  \addcontentsline{toc}{subsection}{Introduction}%
  \label{introduction}%
}




Recently, a unified mathematical framework for the rigorous construction and solution of uncertainty quantification (UQ) problems was formulated {[}\hyperref[oss11]{OSS11}{]}.  This framework, called Optimal Uncertainty Quantification (OUQ), is based on the observation that, given a set of assumptions and information about the problem, there exist optimal bounds on the uncertainties.  These bounds are obtained as extreme values of well-defined optimization problems that correspond to extremizing probabilities of failure subject to the constraints imposed by scenarios compatible with the information set.

An accompanying software framework that implements these rigorous UQ/OUQ methods is now posed.

A rigorous quantification of uncertainty can easily require several thousands of model evaluations $f(x)$. For all but the smallest of models, this requires significant clock time -{}- a model requiring 1 minute of clock time evaluated 10,000 times in a global optimization will take 10,000 minutes ($\sim 7$ days) with a standard optimizer. Furthermore, realistic models are often high-dimensional, highly-constrained, and may require several hours to days even when run on a parallel computer cluster.  For studies of this size or larger to be feasible, a fundamental shift in how we build optimization algorithms is required.  The need to provide support for parallel and distributed computing at the lowest level -{}- within the optimization algorithm -{}- is clear. Standard optimization algorithms must be extended to parallel. The need for new massively-parallel optimization algorithms is also clear.  If these parallel optimizers are not also seamlessly extensible to distributed and heterogeneous computing, then the scope of problems that can be addressed will be severely limited.

While several robust optimization packages exist {[}\hyperref[jop01]{JOP01}, \hyperref[krooo]{KROOO}{]}, there are very few that provide massively-parallel optimization {[}\hyperref[bmm10]{BMM10}, \hyperref[ekl02]{EKL02}, \hyperref[mat09]{MAT09}{]} -{}- the most notable effort being DAKOTA {[}\hyperref[dakot]{DAKOT}{]}, which also includes methods for uncertainty quantification {[}\hyperref[dakuq]{DAKUQ}{]}.
A rethinking of optimization algorithms, from the ground up, is required to dramatically lower the barrier to massively-parallel optimization and rigorous uncertainty quantification. The construction and tight integration of a framework for heterogeneous parallel computing is required to support such optimizations on realistic models. The goal should be to enable widespread availablility of these tools to scientists and engineers in all fields.

Several of the component pieces of such a framework for predictive science already exist, while a few key pieces must be constructed -{}- furthermore, these packages must then be assembled and integrated. Python {[}\hyperref[gvrpy]{GVRPY}{]} is a natural integration environment, and is one that readily supports the dynamic nature of working across heterogeneous resources. By requiring this framework be pure-Python, many of the barriers to running on a new platform are removed. \texttt{multiprocessing} {[}\hyperref[mproc]{MPROC}{]}, \texttt{mpi4py} {[}\hyperref[mpi4p]{MPI4P}{]}, and \texttt{pp} {[}\hyperref[vvppp]{VVPPP}{]} are selected for communication mechanisms, both due to their high level of feature coverage and their relative ease of installation. NumPy {[}\hyperref[numpy]{NUMPY}{]} is used for algorithmic efficiency, and SymPy {[}\hyperref[sym11]{SYM11}{]} is used to provide an alternate interface for building constraints.  Many of the optimization algorithms leverage SciPy {[}\hyperref[jop01]{JOP01}{]}; however like the use of Matplotlib {[}\hyperref[matpl]{MATPL}{]} for plotting, SciPy is an optional dependency.

This paper will discuss the modifications to the \texttt{mystic} {[}\hyperref[mha09]{MHA09}{]} optimization framework required to provide a simple interface to massively parallel optimization, and also to the \texttt{pathos} {[}\hyperref[mba10]{MBA10}{]} framework for staging and launching optimizations on heterogeneous resources.  These efforts leverage \texttt{pyre} {[}\hyperref[maga1]{MAGA1}{]} -{}- an component integration framework for parallel computing, which has recently been extended to distributed communication and management with \texttt{hydra} (part of this development effort). This paper will also overview a new mathematical framework {[}\hyperref[oss11]{OSS11}, \hyperref[all11]{ALL11}, \hyperref[kll11]{KLL11}, \hyperref[loo08]{LOO08}{]} for the quantification of uncertainties, which provides a formulation of UQ problems as global optimization problems.


\subsection*{Rigorous Uncertainty Quantification%
  \phantomsection%
  \addcontentsline{toc}{subsection}{Rigorous Uncertainty Quantification}%
  \label{rigorous-uncertainty-quantification}%
}



Following {[}\hyperref[loo08]{LOO08}{]}, we specifically take a \emph{certification} point of view of uncertainty quantification. For definiteness, we consider systems whose operation can be described in terms of $N$ scalar performance measures $(Y_1,\ldots,Y_N) = Y \in \mathbb{R}^N$. The response of the system is taken as \emph{stochastic} due to the intristic randomness of the system, or randomness in the input parameters defining the operation of the system, or both. Suppose that the outcome $Y \in A$ constitutes a satisfactory outcome for the system of interest, for some prescribed measureable \emph{admissible} set $A \subseteq \mathbb{R}^N$.  Hence, we are interested in determining the \emph{probability of failure} (PoF) $\mathbb{P}[Y \in A^{c}]$.

Evidently, for an upper bound to be useful, it must also be \emph{tight} (i.e. it must be close to the actual PoF of the system) and accessible by some combination of laboratory and computational means. In {[}\hyperref[all11]{ALL11}, \hyperref[kll11]{KLL11}{]}, a methodology for a rigorous determination of tight upper bounds on the probability of failure for complex systems is presented, and is summarized below.

We consider a response function $Y = F(X)$ that maps controllable system inputs $X$ to performance measures $Y$, and relies on a probability of failure (PoF) upper bounds of the concentration of measure (CoM) type {[}\hyperref[bbl04]{BBL04}, \hyperref[led01]{LED01}, \hyperref[mcd89]{MCD89}{]}. If McDiarmid's inequality {[}\hyperref[mcd89]{MCD89}{]} (i.e. the bounded differences inequality) is used to bound PoF, the system may then be certified on the sole knowledge of ranges of its input parameters -{}- without \emph{a priori} knowledge of their probability distributions, its mean performance $\mathbb{E}[Y] = M$ and a certain measure $D_{G} = U$ of the spread of the response, known as \emph{system diameter}, which provides a rigorous quantitative measure of the uncertainty in the response of the system.

A model is regarded as $Y = F(X)$ that approximates the response $Y = G(X)$ of the system. An upper bound on the system diameter -{}- and thus on the uncertainty in the response of the system -{}- then follows from the triangle inequality $D_{G} \leq D_{F} + D_{G-F}$, and $U = D_{F} + D_{G-F}$ can be taken as a new -{}- and conservative -{}- measure of system uncertainty. In this approach, the total uncertainty of the system is the sum of the \emph{predicted uncertainty} (i.e. the variability in performance predicted by the model as quantified by the \emph{model diameter} $D_{F}$), and the \emph{modeling-error uncertainty} (i.e. the discrepancy between model prediction and experiment as quantified by the \emph{modeling-error diameter} $D_{G-F}$.

In {[}\hyperref[loo08]{LOO08}{]}, PoF upper bounds of the CoM type were formulated by recourse to McDiarmid's inequality. In its simplest version, this inequality pertains to a system characterized by $N$ real random inputs $X = (X_1,\ldots,X_N) \in E \subseteq \mathbb{R}^N$ and a single real performance measure $Y \in \mathbb{R}$. Suppose that the function $G : \mathbb{R}^N \to \mathbb{R}$ describes the response function of the system. Suppose that the system fails when $Y \leq a$, where $a$ is a threshold for the safe operation of the system. Then, a direct application of McDiarmid's inequality gives the following upper bound on the PoF of the system:\begin{equation}
\label{eqn-intromcd}
\mathbb{P}[G \leq a]    \leq    \exp\left(-2\frac{M^2}{U^2}\right)
\end{equation}where\begin{equation}
\label{eqn-introm}
M = (\mathbb{E}[G]-a)_+
\end{equation}is the \emph{design margin} and\begin{equation}
\label{eqn-introu}
U = D_{G}
\end{equation}
is the \emph{system uncertainty}. In (\DUrole{ref}{eqn-introu}), $D_{G}$ is the diameter of the response function. From (\DUrole{ref}{eqn-intromcd}) it follows that the system is certified if\begin{equation*}
\exp\left(-2\frac{M^2}{U^2}\right) \leq \epsilon
\end{equation*}where $\epsilon$ is the PoF tolerance, or, equivalently, if\begin{equation}
\label{eqn-introcf}
\text{CF} = \frac{M}{U} \geq \sqrt{\log\sqrt{\frac{1}{\epsilon}}}
\end{equation}where $\text{CF}$ is the \emph{confidence factor}. In writing (\DUrole{ref}{eqn-introm}) and subsequently, we use the function $x_+:=\max(0,x)$. We see from the preceding expressions that McDiarmid's inequality supplies rigorous quantitative definitions of design margin and system uncertainty. In particular, the latter is measured by \emph{system diameter} $D_G$, which measures the largest deviation in performance resulting from arbitrarily large perturbations of one input parameter at a time. Within this simple framework, rigorous certification is achieved by the determination of two-{}-and only two-{}-quantities: the \emph{mean performance} $\mathbb{E}[G]$ and the \emph{system diameter} $D_G$.


McDiarmid's inequality is a result in probability theory that provides an upper bound on the probability that the value of a function depending on multiple independent random variables deviates from its expected value. A central device in McDiarmid's inequality is the \emph{diameter} of a function. We begin by recalling that the \emph{oscillation} $\operatorname{osc}(f,E)$ of a real function $f : E \to \mathbb{R}$ over a set $E \in R$ is\begin{equation}
\label{eqn-introosc}
\operatorname{osc}(f,E) = \sup \{|f(y) - f(x)| \,:\, x,y \in E \}
\end{equation}Thus, $\operatorname{osc}(f,E)$ measures the spread of values of $f$ that may be obtained by allowing the independent variables to range over its entire domain of definition. For functions $f : E \subset \mathbb{R}^N \to \mathbb{R}$ of several real values, component-wise  \emph{suboscillations} can be defined as\begin{equation}
\label{eqn-subosc}
\operatorname{osc}_i(f,E) = \sup \{|f(y) - f(x)| \,:\, x,y \in E,\,\, x_{j} = y_{j} \,\,\,\text{for}\,\,\, j \neq i \}
\end{equation}Thus $\operatorname{osc}_i(f,E)$ measures the maximum oscillation among all one-dimensional fibers in the direction of the $i\text{th}$ coordinate. The \emph{diameter} $D(f,E)$ of the function $f : E \to \mathbb{R}$ is obtained as the root-mean square of its component-wise suboscillations:\begin{equation}
\label{eqn-diamosc}
D(f,E) = \left( \sum_{i=1}^{n} \operatorname{osc}_{i}^{2}(f,E) \right)^{1/2}
\end{equation}and it provides a measure of the spread of the range of the function.
Thus (\DUrole{ref}{eqn-subosc}) also us to regard $\operatorname{osc}_i(f,E)$ as a \emph{subdiameter} of the system corresponding to variable $X_{i}$, where the subdiameter can be regarded as a measure of uncertainty contributed by the variable $X_{i}$ to the total uncertainty of the system.


The attractiveness of the McDiarmid CoM approach to UQ relies on the requirement of tractable information on response functions (sub-diameters) and measures (independence and mean response). Above, it is described how to ``plug'' this information into McDiarmid's concentration inequality to obtain an upper bound on probabilies of deviation. One may wonder if it is possible to obtain an ``optimal'' concentration inequality, especially when the available information may not necessarily be sub-diameters and mean values. A general mathematical framework for optimally quantifying uncertainties based only on available information has been proposed {[}\hyperref[oss11]{OSS11}{]}, and will be summarized here. Assume, for instance, that one wants to certify that\begin{equation}
\label{eqn-ouqpof}
\mathbb{P}[G \geq a]    \leq    \epsilon
\end{equation}based on the information that $\operatorname{osc}_i(G,E) \leq D_{i}$, $X = (X_{1},\ldots,X_{N})$, $\mathbb{E}[G] \leq 0$ and that the inputs $X_{i}$ are independent under $\mathbb{P}$. In this situation, the optimal upper bound $\mathcal{U}(\mathcal{A}_{MD})$ on the PoF $\mathbb{P}[G \geq a]$ is the solution of the following optimization problem
\begin{equation}
\label{eqn-ouqupper}
\mathcal{U}(\mathcal{A}_{MD}) = \sup_{(f,\mu)\in\mathcal{A}_{MD}} \mu[f(X) \geq a]
\end{equation}subject to constraints provied by the information set\begin{equation}
\label{eqn-ouqamcd}
\mathcal{A}_{MD} = \left\{ (f, \mu) \,\middle|\,
    \begin{matrix}
        f \,:\, E_{1} \times \dots \times E_{N} \to \mathbb{R}, \\
        \mu \in \mathcal{M}(E_{1}) \otimes \dots \otimes \mathcal{M}(E_{N}), \\
        \mathbb{E}_{\mu}[f] \leq 0, \\
        \operatorname{osc}_{i}(f,E) \leq D_{i}
    \end{matrix} \right\}
\end{equation}where $\mathcal{M}(E_{k})$ denotes the set of measures of probability on $E_{k}$. Hence, McDiarmid's inequality is the statement that\begin{equation}
\label{eqn-ouqmcdsoln}
\mathcal{U}(\mathcal{A}_{MD})    \leq    \exp\left(-2\frac{a^2}{\sum_{i=1}^{N} D_{i}^2}\right)
\end{equation}Similarly, for any other set of information $\mathcal{A}$, we have an optimal (i.e.) least upper bound on the probability of deviation\begin{equation}
\label{eqn-ouqgeneral}
\mathcal{U}(\mathcal{A}) = \sup_{(f,\mu)\in\mathcal{A}} \mu[f(X) \geq a]
\end{equation}The idea is that in practical applications, the available information does not determine $(G,\mathbb{P})$ uniquely, but does determine a set $\mathcal{A}$ such that $(G,\mathbb{P}) \in \mathcal{A}$ and such that any $(f,\mathbb{\mu}) \in \mathcal{A}$ could \emph{a priori} be $(G,\mathbb{P})$. This mathematical framework, called optimal uncertainty quantification (OUQ), is based on the observation that, given a set of assumptions and information about the problem, there exist optimal bounds on uncertainties; these are obtained as extreme values of well-defined optimization problems corresponding to extremizing probabilities of failure, or of deviations, over the feasible set $\mathcal{A}$. Observe that this framework does not implicitly impose inappropriate assumptions, nor does it repudiate relevant information. Indeed, as demonstrated in (\DUrole{ref}{eqn-ouqamcd} and \DUrole{ref}{eqn-ouqmcdsoln}) for the CoM approach, OUQ can pose a problem that incorporates the assumptions utilized in other common UQ methods (such as Bayesian inference {[}\hyperref[ljh99]{LJH99}{]}) and provide a rigorous optimal bound on the uncertainties.


Although some OUQ problems can be solved analytically, most must be solved numerically.  To that end, the reduction theorems of {[}\hyperref[oss11]{OSS11}{]} reduce the infinite-dimensional feasible set $\mathcal{A}$ to a finite-dimensional subset $\mathcal{A}_{\Delta}$ that has the key property that the objective function (PoF) has the same lower and upper extreme values over $\mathcal{A}_{\Delta}$ as over $\mathcal{A}$.

For example, the reduction for $\mathcal{A}_{MD}$ in (\DUrole{ref}{eqn-ouqamcd}) is to pass to measures $\mu = \mu_{1} \otimes \dots \otimes \mu_{N}$ such that each marginal measure $\mu_{i}$ is supported on at most two points of the parameter space $E_{i}$, i.e. $\mu_{i}$ is a convex combination of two Dirac measures (point masses).  Having reduced the set of feasible measures $\mu$, the set of feasible response functions $f$ is also reduced, since we only care about the values of $f$ on the finite support of $\mu$ and nowhere else.

We refer the reader to {[}\hyperref[oss11]{OSS11}{]} for the more general reduction theorems.  The essential point is that if the information/constraints take the form of $n_{i}$ inequalities of the form $\mathbb{E}_{\mu_{i}}[\phi_{j}] \leq 0$ (for some test functions $\phi_{j}$) and $n'$ inequalities of the form $\mathbb{E}_{\mu}[\phi_{j}] \leq 0$, then it is enough to consider $\mu_{i}$ with support on $1 + n_{i} + n'$ points of $E_{i}$.

The reduction theorems leave us with a finite-dimensional optimization problem in which the optimization variables are suitable parametrizations of the \emph{reduced} feasible scenarios $(f, \mu)$.


\subsection*{A Highly-Configurable Optimization Framework%
  \phantomsection%
  \addcontentsline{toc}{subsection}{A Highly-Configurable Optimization Framework}%
  \label{a-highly-configurable-optimization-framework}%
}

We have built a robust optimization framework (\texttt{mystic}) {[}\hyperref[mha09]{MHA09}{]} that incorporates the mathematical framework described in {[}\hyperref[oss11]{OSS11}{]}, and have provided an interface to prediction, certification, and validation as a framework service. The \texttt{mystic} framework provides a collection of optimization algorithms and tools that lowers the barrier to solving complex optimization problems. \texttt{mystic} provides a selection of optimizers, both global and local, including several gradient solvers.  A unique and powerful feature of the framework is the ability to apply and configure solver-independent termination conditions -{}-{}- a capability that greatly increases the flexibility for numerically solving problems with non-standard convergence profiles. All of \texttt{mystic}'s solvers conform to a solver API, thus also have common method calls to configure and launch an optimization job. This allows any of \texttt{mystic}'s solvers to be easily swapped without the user having to write any new code.

The minimal solver interface:\begin{Verbatim}[commandchars=\\\{\},fontsize=\footnotesize]
\PY{c}{\PYZsh{} the function to be minimized and the initial values}
\PY{k+kn}{from} \PY{n+nn}{mystic.models} \PY{k+kn}{import} \PY{n}{rosen} \PY{k}{as} \PY{n}{my\PYZus{}model}
\PY{n}{x0} \PY{o}{=} \PY{p}{[}\PY{l+m+mf}{0.8}\PY{p}{,} \PY{l+m+mf}{1.2}\PY{p}{,} \PY{l+m+mf}{0.7}\PY{p}{]}

\PY{c}{\PYZsh{} configure the solver and obtain the solution}
\PY{k+kn}{from} \PY{n+nn}{mystic.solvers} \PY{k+kn}{import} \PY{n}{fmin}
\PY{n}{solution} \PY{o}{=} \PY{n}{fmin}\PY{p}{(}\PY{n}{my\PYZus{}model}\PY{p}{,} \PY{n}{x0}\PY{p}{)}
\end{Verbatim}
\begin{figure}[]
\noindent{\includegraphics[scale=0.300000]{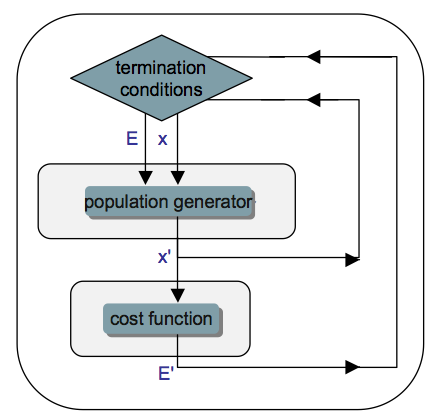}\hfill}
\caption{Conceptual diagram for an optimizer. The cost function provides a difference metric that accepts input parameters $x$ and produces a cost $E$. \DUrole{label}{fig-optimization}}
\end{figure}
The criteria for when and how an optimization terminates are of paramount importance in traversing a function's potential well. Standard optimization packages provide a single convergence condition for each optimizer. \texttt{mystic} provides a set of fully customizable termination conditions, allowing the user to discover how to better navigate the optimizer through difficult terrain. Optimizers can be further configured through several methods for choosing the \texttt{InitialPoints}.

The expanded solver interface:\begin{Verbatim}[commandchars=\\\{\},fontsize=\footnotesize]
\PY{c}{\PYZsh{} the function to be minimized and initial values}
\PY{k+kn}{from} \PY{n+nn}{mystic.models} \PY{k+kn}{import} \PY{n}{rosen} \PY{k}{as} \PY{n}{my\PYZus{}model}
\PY{n}{x0} \PY{o}{=} \PY{p}{[}\PY{l+m+mf}{0.8}\PY{p}{,} \PY{l+m+mf}{1.2}\PY{p}{,} \PY{l+m+mf}{0.7}\PY{p}{]}

\PY{c}{\PYZsh{} get monitor and termination condition objects}
\PY{k+kn}{from} \PY{n+nn}{mystic.monitors} \PY{k+kn}{import} \PY{n}{Monitor}\PY{p}{,} \PY{n}{VerboseMonitor}
\PY{n}{stepmon} \PY{o}{=} \PY{n}{VerboseMonitor}\PY{p}{(}\PY{l+m+mi}{5}\PY{p}{)}
\PY{n}{evalmon} \PY{o}{=} \PY{n}{Monitor}\PY{p}{(}\PY{p}{)}
\PY{k+kn}{from} \PY{n+nn}{mystic.termination} \PY{k+kn}{import} \PY{n}{ChangeOverGeneration}
\PY{n}{COG} \PY{o}{=} \PY{n}{ChangeOverGeneration}\PY{p}{(}\PY{p}{)}

\PY{c}{\PYZsh{} instantiate and configure the solver}
\PY{k+kn}{from} \PY{n+nn}{mystic.solvers} \PY{k+kn}{import} \PY{n}{NelderMeadSimplexSolver}
\PY{n}{solver} \PY{o}{=} \PY{n}{NelderMeadSimplexSolver}\PY{p}{(}\PY{n+nb}{len}\PY{p}{(}\PY{n}{x0}\PY{p}{)}\PY{p}{)}
\PY{n}{solver}\PY{o}{.}\PY{n}{SetInitialPoints}\PY{p}{(}\PY{n}{x0}\PY{p}{)}
\PY{n}{solver}\PY{o}{.}\PY{n}{SetGenerationMonitor}\PY{p}{(}\PY{n}{stepmon}\PY{p}{)}
\PY{n}{solver}\PY{o}{.}\PY{n}{SetEvaluationMonitor}\PY{p}{(}\PY{n}{evalmon}\PY{p}{)}
\PY{n}{solver}\PY{o}{.}\PY{n}{Solve}\PY{p}{(}\PY{n}{my\PYZus{}model}\PY{p}{,} \PY{n}{COG}\PY{p}{)}

\PY{c}{\PYZsh{} obtain the solution}
\PY{n}{solution} \PY{o}{=} \PY{n}{solver}\PY{o}{.}\PY{n}{bestSolution}

\PY{c}{\PYZsh{} obtain diagnostic information}
\PY{n}{function\PYZus{}evals} \PY{o}{=} \PY{n}{solver}\PY{o}{.}\PY{n}{evaluations}
\PY{n}{iterations} \PY{o}{=} \PY{n}{solver}\PY{o}{.}\PY{n}{generations}
\PY{n}{cost} \PY{o}{=} \PY{n}{solver}\PY{o}{.}\PY{n}{bestEnergy}

\PY{c}{\PYZsh{} modify the solver configuration, and continue}
\PY{n}{COG} \PY{o}{=} \PY{n}{ChangeOverGeneration}\PY{p}{(}\PY{n}{tolerance}\PY{o}{=}\PY{l+m+mf}{1e-8}\PY{p}{)}
\PY{n}{solver}\PY{o}{.}\PY{n}{Solve}\PY{p}{(}\PY{n}{my\PYZus{}model}\PY{p}{,} \PY{n}{COG}\PY{p}{)}

\PY{c}{\PYZsh{} obtain the new solution}
\PY{n}{solution} \PY{o}{=} \PY{n}{solver}\PY{o}{.}\PY{n}{bestSolution}
\end{Verbatim}


\texttt{mystic} provides progress monitors that can be attached to an optimizer to track progress of the fitted parameters and the value of the cost function.
Additionally, monitors can be customized to track the function gradient or other progress metrics. Monitors can also be configured to record either function evaluations or optimization iterations (i.e. \emph{generations}).
For example, using \texttt{VerboseMonitor(5)} in the \texttt{SetGenerationMonitor} method will print the \texttt{bestEnergy} to \texttt{stdout} every five generations.\begin{figure}[]
\noindent{\includegraphics[width=\columnwidth]{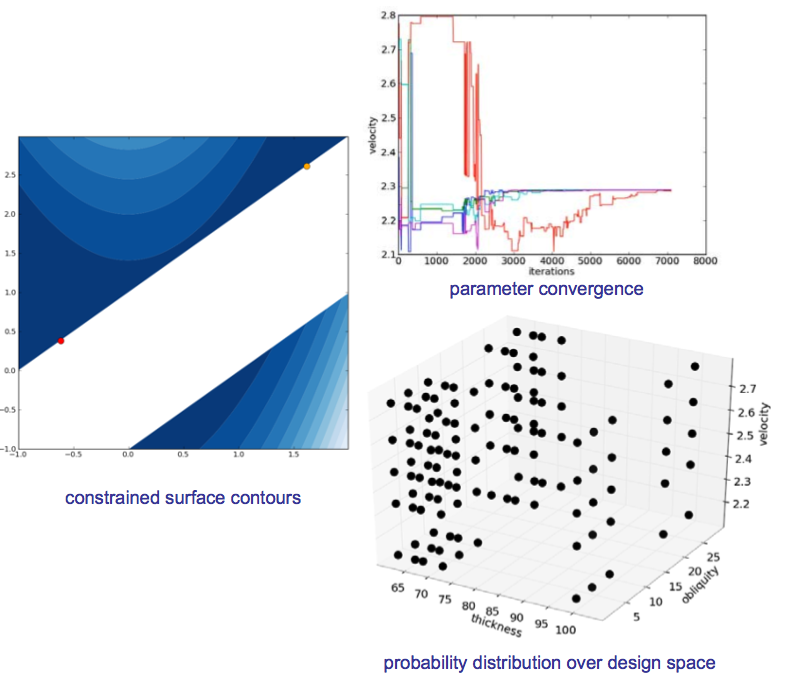}\hfill}
\caption{Optimization analysis viewers available in \texttt{mystic}. \DUrole{label}{fig-viewers}}
\end{figure}
\begin{figure}[]
\noindent{\includegraphics[scale=0.350000]{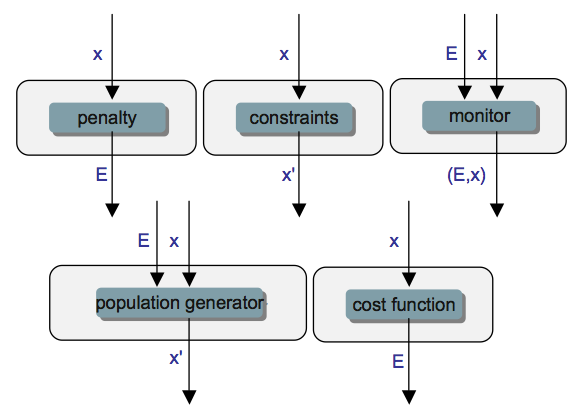}\hfill}
\caption{Basic components provided in the optimizer toolkit. Several wrapper classes are also provided for binding components, while factory classes are provided for generating components. \DUrole{label}{fig-custom}}
\end{figure}



\subsection*{Constraints Toolkit%
  \phantomsection%
  \addcontentsline{toc}{subsection}{Constraints Toolkit}%
  \label{constraints-toolkit}%
}


\texttt{mystic} provides a method to constrain optimization to be within an $N$-dimensional box on input space, and also a method to impose user-defined parameter constraint functions on any cost function.  Thus, both \emph{bounds constraints} and \emph{parameter constraints} can be generically applied to any of \texttt{mystic}'s unconstrained optimization algorithms. Traditionally, constrained optimization problems tend to be solved iteratively, where a penalty is applied to candidate solutions that violate the constraints. Decoupling the solving of constraints from the optimization problem can greatly increase the efficiency in solving highly-constrained nonlinear problems -{}- effectively, the optimization algorithm only selects points that satisfy the constraints. Constraints can be solved numerically or algebraically, where the solving of constraints can itself be cast as an optimization problem. Constraints can also be dynamically applied, thus altering an optimization in progress.

Penalty-based methods indirectly modify the candidate solution by applying a change in energy $\Delta E = k \cdot p(\vec{x})$ to the unconstrained cost function $f(\vec{x})$ when the constraints are violated. The modified cost function $\phi$ is thus written as:\begin{equation}
\label{eqn-penaltycon}
\phi(\vec{x}) = f(\vec{x}) + k \cdot p(\vec{x})
\end{equation}Set-based methods directly modify the candidate solution by applying a constraints solver $c$ that ensures the optimizer will always select from a set of candidates that satisfy the constraints. The constraints solver has an interface ${\vec{x}\,}' = c(\vec{x})$, and the cost function becomes:\begin{equation}
\label{eqn-directcon}
\phi(\vec{x}) = f(c(\vec{x}))
\end{equation}Adding parameter constraints to a solver is as simple as building a constraints function, and using the \texttt{SetConstraints} method. Additionally, simple bounds constraints can also be applied through the \texttt{SetStrictRanges} method:\begin{Verbatim}[commandchars=\\\{\},fontsize=\footnotesize]
\PY{c}{\PYZsh{} a user-provided constraints function}
\PY{k}{def} \PY{n+nf}{constrain}\PY{p}{(}\PY{n}{x}\PY{p}{)}\PY{p}{:}
  \PY{n}{x}\PY{p}{[}\PY{l+m+mi}{1}\PY{p}{]} \PY{o}{=} \PY{n}{x}\PY{p}{[}\PY{l+m+mi}{0}\PY{p}{]}
  \PY{k}{return} \PY{n}{x}

\PY{c}{\PYZsh{} the function to be minimized and the bounds}
\PY{k+kn}{from} \PY{n+nn}{mystic.models} \PY{k+kn}{import} \PY{n}{rosen} \PY{k}{as} \PY{n}{my\PYZus{}model}
\PY{n}{lb} \PY{o}{=} \PY{p}{[}\PY{l+m+mf}{0.0}\PY{p}{,} \PY{l+m+mf}{0.0}\PY{p}{,} \PY{l+m+mf}{0.0}\PY{p}{]}
\PY{n}{ub} \PY{o}{=} \PY{p}{[}\PY{l+m+mf}{2.0}\PY{p}{,} \PY{l+m+mf}{2.0}\PY{p}{,} \PY{l+m+mf}{2.0}\PY{p}{]}

\PY{c}{\PYZsh{} get termination condition object}
\PY{k+kn}{from} \PY{n+nn}{mystic.termination} \PY{k+kn}{import} \PY{n}{ChangeOverGeneration}
\PY{n}{COG} \PY{o}{=} \PY{n}{ChangeOverGeneration}\PY{p}{(}\PY{p}{)}

\PY{c}{\PYZsh{} instantiate and configure the solver}
\PY{k+kn}{from} \PY{n+nn}{mystic.solvers} \PY{k+kn}{import} \PY{n}{NelderMeadSimplexSolver}
\PY{n}{solver} \PY{o}{=} \PY{n}{NelderMeadSimplexSolver}\PY{p}{(}\PY{n+nb}{len}\PY{p}{(}\PY{n}{x0}\PY{p}{)}\PY{p}{)}
\PY{n}{solver}\PY{o}{.}\PY{n}{SetRandomInitialPoints}\PY{p}{(}\PY{n}{lb}\PY{p}{,} \PY{n}{ub}\PY{p}{)}
\PY{n}{solver}\PY{o}{.}\PY{n}{SetStrictRanges}\PY{p}{(}\PY{n}{lb}\PY{p}{,} \PY{n}{ub}\PY{p}{)}
\PY{n}{solver}\PY{o}{.}\PY{n}{SetConstraints}\PY{p}{(}\PY{n}{constrain}\PY{p}{)}
\PY{n}{solver}\PY{o}{.}\PY{n}{Solve}\PY{p}{(}\PY{n}{my\PYZus{}model}\PY{p}{,} \PY{n}{COG}\PY{p}{)}

\PY{c}{\PYZsh{} obtain the solution}
\PY{n}{solution} \PY{o}{=} \PY{n}{solver}\PY{o}{.}\PY{n}{bestSolution}
\end{Verbatim}

\texttt{mystic} provides a simple interface to a lot of underlying complexity -{}- thus allowing a non-specialist user to easily access optimizer configurability and high-performance computing without a steep learning curve. This feature must also be applied to the application of constraints on a function or measure. The natural syntax for a constraint is one of symbolic math, hence \texttt{mystic} leverages SymPy {[}\hyperref[sym11]{SYM11}{]} to construct a symbolic math parser for the translation of the user's input into functioning constraint code objects:\begin{Verbatim}[commandchars=\\\{\},fontsize=\footnotesize]
\PY{c}{\PYZsh{} a user-provided constraints function}
\PY{n}{constraints} \PY{o}{=} \PY{l+s}{"""}
\PY{l+s}{x2 = x1}
\PY{l+s}{"""}
\PY{k+kn}{from} \PY{n+nn}{mystic.constraints} \PY{k+kn}{import} \PY{n}{parse}
\PY{n}{constrain} \PY{o}{=} \PY{n}{parse}\PY{p}{(}\PY{n}{constraints}\PY{p}{)}
\end{Verbatim}
The constraints parser is a constraints factory method that can parse multiple and nonlinear constraints, hard or soft (i.e. ``$\sim$'') constraints, and equality or inequality (i.e. ``$>$'') constraints.

Similar tools exist for creating penalty functions, including a \texttt{SetPenalty} method for solvers. Available penalty methods include the exterior penalty function method {[}\hyperref[ven09]{VEN09}{]}, the augmented Lagrange multiplier method {[}\hyperref[ksk94]{KSK94}{]}, and the logarithmic barrier method {[}\hyperref[jjb03]{JJB03}{]}. At the low-level, penalty functions are bound to the cost function using \texttt{mystic}'s \texttt{functionWrapper} method.

It is worth noting that the use of a constraints solver $c$ does not require the constraints be bound to the cost function. The evaluation of the constraints are decoupled from the evaluation of the cost function -{}- hence, with \texttt{mystic}, highly-constrained optimization decomposes to the solving of $K$ independent constraints, followed by an unconstrained optimization over only the set of valid points. This method has been shown effective for solving optimization problems where $K \approx 200$ {[}\hyperref[oss11]{OSS11}{]}.


\subsection*{Seamless Migration to Parallel Computing%
  \phantomsection%
  \addcontentsline{toc}{subsection}{Seamless Migration to Parallel Computing}%
  \label{seamless-migration-to-parallel-computing}%
}

\texttt{mystic} is built from the ground up to utilize parallel and distributed computing. The decomposition of optimization algorithms into their component parts allow this decomposition to not only be in an abstraction layer, but across process-space. \texttt{mystic} provides a \texttt{modelFactory} method that convers a user's model to a \emph{service}. We define a service to be an entity that is callable by globally unique identifier. Services can also be called by proxy. In \texttt{mystic}, services also include infrastructure for monitoring and handling events. An optimization is then composed as a network of interacting services, with the most common being the user's model or cost function being mapped over parallel resources.

\texttt{mystic} provides several stock models and model factories that are useful for testing:\begin{Verbatim}[commandchars=\\\{\},fontsize=\footnotesize]
\PY{c}{\PYZsh{} generate a model from a stock 'model factory'}
\PY{k+kn}{from} \PY{n+nn}{mystic.models.lorentzian} \PY{k+kn}{import} \PY{n}{Lorentzian}
\PY{n}{lorentz} \PY{o}{=} \PY{n}{Lorentzian}\PY{p}{(}\PY{n}{coeffs}\PY{p}{)}

\PY{c}{\PYZsh{} evaluate the model}
\PY{n}{y} \PY{o}{=} \PY{n}{lorentz}\PY{p}{(}\PY{n}{x}\PY{p}{)}
\end{Verbatim}
\begin{figure}[]
\noindent{\includegraphics[scale=0.300000]{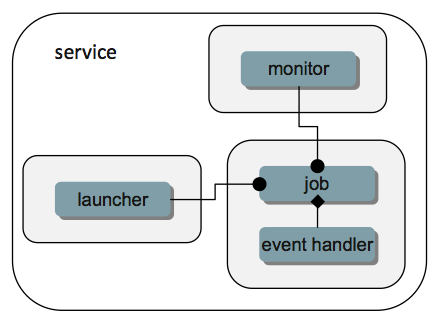}\hfill}
\caption{Conceptual diagram for a service-based model. Here, the job is the fundamental commodity of work, and is the object on which the service is based -{}- in \texttt{mystic}, this is typically the user's model or a cost function. Services have a global unique identifier, and thus can easily be called by proxy. Note that services may not be located on the machine that requested the service be spawned. Services also can be imbued with infrastructure for monitoring and handling events. Monitors write to a stream that can be piped into another object, such as a logger or one of \texttt{mystic}'s viewers. \DUrole{label}{fig-service}}
\end{figure}
Model factory methods insert \texttt{pathos} infrastructure, thus casting a model as a callable \emph{service} that has been imbued with \texttt{pathos} infrastructure as shown in Figure (\DUrole{ref}{fig-service}). The default \texttt{launcher} and \texttt{map} included in \texttt{mystic} are functionally equivalent to execution and \texttt{map} within the standard Python distribution.  Any user-provided function can be cast as a service through the use of a \texttt{modelFactory}:\begin{Verbatim}[commandchars=\\\{\},fontsize=\footnotesize]
\PY{c}{\PYZsh{} a user-provided model function}
\PY{k}{def} \PY{n+nf}{identify}\PY{p}{(}\PY{n}{x}\PY{p}{)}
  \PY{k}{return} \PY{n}{x}

\PY{c}{\PYZsh{} add pathos infrastructure (included in mystic)}
\PY{k+kn}{from} \PY{n+nn}{mystic.tools} \PY{k+kn}{import} \PY{n}{modelFactory}\PY{p}{,} \PY{n}{Monitor}
\PY{n}{evalmon} \PY{o}{=} \PY{n}{Monitor}\PY{p}{(}\PY{p}{)}
\PY{n}{my\PYZus{}model} \PY{o}{=} \PY{n}{modelFactory}\PY{p}{(}\PY{n}{identify}\PY{p}{,} \PY{n}{monitor}\PY{o}{=}\PY{n}{evalmon}\PY{p}{)}

\PY{c}{\PYZsh{} evaluate the model}
\PY{n}{y} \PY{o}{=} \PY{n}{my\PYZus{}model}\PY{p}{(}\PY{n}{x}\PY{p}{)}

\PY{c}{\PYZsh{} evaluate the model with a map function}
\PY{k+kn}{from} \PY{n+nn}{mystic.tools} \PY{k+kn}{import} \PY{n}{PythonMap}
\PY{n}{my\PYZus{}map} \PY{o}{=} \PY{n}{PythonMap}\PY{p}{(}\PY{p}{)}
\PY{n}{z} \PY{o}{=} \PY{n}{my\PYZus{}map}\PY{p}{(}\PY{n}{my\PYZus{}model}\PY{p}{,} \PY{n+nb}{range}\PY{p}{(}\PY{l+m+mi}{10}\PY{p}{)}\PY{p}{)}
\end{Verbatim}


\begin{figure}[]
\noindent{\includegraphics[width=\columnwidth]{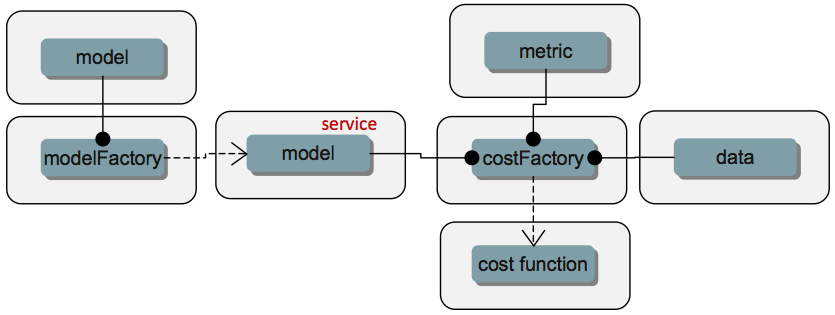}\hfill}
\caption{Use of a \texttt{modelFactory} to cast a user's model $F(x)$ as a service. The model and experimental data $G$ are then bound with a \texttt{costFactory} to produce a cost function. A \texttt{costFactory} can accept a raw user's model, a model proxy, or a model service (as shown here). A typical metric is $|F(x) - G|^{2}$.  \DUrole{label}{fig-modelfactory}}
\end{figure}


\subsection*{A Framework for Heterogeneous Computing%
  \phantomsection%
  \addcontentsline{toc}{subsection}{A Framework for Heterogeneous Computing}%
  \label{a-framework-for-heterogeneous-computing}%
}
We have developed a computational job management framework (\texttt{pathos}) {[}\hyperref[mba10]{MBA10}{]} that offers a simple, efficient, and consistent user experience in a variety of heterogeneous environments from multi-core workstations to networks of large-scale computer clusters. \texttt{pathos} provides a single environment for developing and testing algorithms locally -{}- and enables the user to execute the algorithms on remote clusters, while providing the user with full access to their job history. \texttt{pathos} primarily provides the communication mechanisms for configuring and launching parallel computations across heterogenous resources. \texttt{pathos} provides stagers and launchers for parallel and distributed computing, where each launcher contains the syntactic logic to configure and launch jobs in an execution environment. Some examples of included launchers are: a queue-less MPI-based launcher, a SSH-based launcher, and a \texttt{multiprocessing} launcher. \texttt{pathos} also provides a map-reduce algorithm for each of the available launchers, thus greatly lowering the barrier for users to extend their code to parallel and distributed resources. \texttt{pathos} provides the ability to interact with batch schedulers and queuing systems, thus allowing large computations to be easily launched on high-performance computing resources. One of the most powerful features of \texttt{pathos} is \texttt{sshTunnel}, which enables a user to automatically wrap any distributed service calls within an SSH tunnel.

\texttt{pathos} is divided into four subpackages: \texttt{dill} (a utility for serialization of Python objects), \texttt{pox} (utilities for filesystem exploration and automated builds), \texttt{pyina} (a MPI-based parallel mapper and launcher), and \texttt{pathos} (distributed parallel map-reduce and SSH communication).

\texttt{pathos} utilizes \texttt{pyre}, which provides tools for connecting components and managing their interactions. The core component used by \texttt{pathos} is a service -{}- a callable object with a configurable connection mechanism. A service can utilize \texttt{Launcher} and \texttt{Monitor} objects (which provide abstraction to execution and logging, respectively), as well as \texttt{Strategy} objects (which provide abstraction patterns for coupling services). A standard interface for services enables massively parallel applications that utilize distributed resources to be constructed from a few simple building blocks.  A \texttt{Launcher} contains the logic required to initiate execution on the current execution environment. The selection of launcher will determine if the code is submitted to a batch queue, run across SSH tunneled RPC connections, or run with MPI on a multiprocessor.  A \texttt{Strategy} provides an algorithm to distribute the workload among available resources. Strategies can be static or dynamic. Examples of static strategies include the \texttt{equalportion} strategy and the \texttt{carddealer} strategy. Dynamic strategies are based on the concept of a worker \texttt{pool}, where there are several workload balancing options to choose from.  Strategies and launchers can be coupled together to provide higher-level batch and parallel-map algorithms. A \texttt{Map} interface allows batch processing to be decoupled from code execution details on the selected platforms, thus enabling the same application to be utilized for sequential, parallel, and distributed parallel calculations.


\subsection*{Globally Unique Message Passing%
  \phantomsection%
  \addcontentsline{toc}{subsection}{Globally Unique Message Passing}%
  \label{globally-unique-message-passing}%
}

We must design for the case where an optimizer's calculation spans multiple clusters, with a longevity that may exceed the uptime of any single cluster or node.  \texttt{hydra} enables any Python object to obtain a network address. After obtaining an address, an object can asynchronously exchange messages with other objects on the network.  Through the use of proxy objects, sending messages to remote objects is easy as calling an instance method on a local object.  A call to a proxy transparently pickles the function name along with the arguments, packages the message as a datagram, and sends it over the network to the remote object represented by the proxy.  On the recieving end, there is a mechanism for responding to the sender of the current message. Since message sending is asynchronous, an object responds to a message by sending another message.

The \texttt{modelFactory} method essentially provides \texttt{mystic} with a high-level interface for a \texttt{pathos} server, with an option to bind a monitor directly to the service.  The lower-level construction of a distributed service, using SSH-based communication, is as follows:\begin{Verbatim}[commandchars=\\\{\},fontsize=\footnotesize]
\PY{c}{\PYZsh{} a user-provided model function}
\PY{k}{def} \PY{n+nf}{identify}\PY{p}{(}\PY{n}{x}\PY{p}{)}
  \PY{k}{return} \PY{n}{x}

\PY{c}{\PYZsh{} cast the model as a distributed service}
\PY{k+kn}{from} \PY{n+nn}{pathos.servers} \PY{k+kn}{import} \PY{n}{sshServer}
\PY{n+nb}{id} \PY{o}{=} \PY{l+s}{'}\PY{l+s}{foo.caltech.edu:50000:spike42}\PY{l+s}{'}
\PY{n}{my\PYZus{}proxy} \PY{o}{=} \PY{n}{sshServer}\PY{p}{(}\PY{n}{identify}\PY{p}{,} \PY{n}{server}\PY{o}{=}\PY{n+nb}{id}\PY{p}{)}

\PY{c}{\PYZsh{} evaluate the model via proxy}
\PY{n}{y} \PY{o}{=} \PY{n}{my\PYZus{}proxy}\PY{p}{(}\PY{n}{x}\PY{p}{)}
\end{Verbatim}
Parallel map functions are built around available launchers, providing a high-level interface to launching several copies of a model in parallel. The creation of a parallel map that will draw from a pool of two local workers and all available IPC servers at \texttt{'foo.caltech.edu'} is shown below:\begin{Verbatim}[commandchars=\\\{\},fontsize=\footnotesize]
\PY{c}{\PYZsh{} a user-provided model function}
\PY{k}{def} \PY{n+nf}{identify}\PY{p}{(}\PY{n}{x}\PY{p}{)}
  \PY{k}{return} \PY{n}{x}

\PY{c}{\PYZsh{} select and configure a parallel map}
\PY{k+kn}{from} \PY{n+nn}{pathos.maps} \PY{k+kn}{import} \PY{n}{ipcPool}
\PY{n}{my\PYZus{}map} \PY{o}{=} \PY{n}{ipcPool}\PY{p}{(}\PY{l+m+mi}{2}\PY{p}{,} \PY{n}{servers}\PY{o}{=}\PY{p}{[}\PY{l+s}{'}\PY{l+s}{foo.caltech.edu}\PY{l+s}{'}\PY{p}{]}\PY{p}{)}

\PY{c}{\PYZsh{} evaluate the model in parallel}
\PY{n}{z} \PY{o}{=} \PY{n}{my\PYZus{}map}\PY{p}{(}\PY{n}{identify}\PY{p}{,} \PY{n+nb}{range}\PY{p}{(}\PY{l+m+mi}{10}\PY{p}{)}\PY{p}{)}
\end{Verbatim}


\begin{figure}[]
\noindent{\includegraphics[width=\columnwidth]{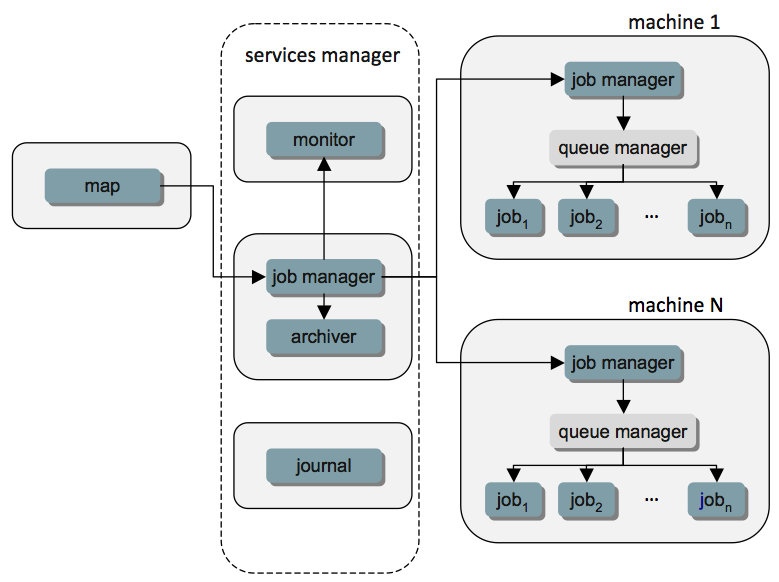}\hfill}
\caption{Conceptual diagram for heterogeneous job management. A distributed parallel map function is used to copy a service $n$ times on $N$ machines. If the object being mapped is not a service, then the services manager is omitted from the diagram -{}- the jobs still undergo a distributed launch, but are managed at the machine level. \DUrole{label}{fig-jobmanagement}}
\end{figure}


\subsection*{Serialization%
  \phantomsection%
  \addcontentsline{toc}{subsection}{Serialization}%
  \label{serialization}%
}
\texttt{dill} extends Python's \texttt{pickle} module for serializing and de-serializing Python objects to the majority of the built-in Python and NumPy types. Serialization is the process of converting an object to a byte stream, the inverse of which is converting a byte stream back to a Python object hierarchy.

\texttt{dill} provides the user the same interface as the \texttt{pickle} module, and also includes some additional features. In addition to pickling Python objects, \texttt{dill} provides the ability to save the state of an interpreter session in a single command. Hence, it would be feasible to save a interpreter session, close the interpreter, ship the pickled file to another computer, open a new interpreter, unpickle the session and thus continue from the ``saved'' state of the original interpreter session.


\subsection*{Filesystem Interaction%
  \phantomsection%
  \addcontentsline{toc}{subsection}{Filesystem Interaction}%
  \label{filesystem-interaction}%
}

\texttt{pox} provides a collection of utilities for navigating and manipulating filesystems. This module is designed to facilitate some of the low level operating system interactions that are useful when exploring a filesystem on a remote host, where queries such as ``what is the root of the filesystem?'', ``what is the user's name?'', and ``what login shell is preferred?'' become essential in allowing a remote user to function as if they were logged in locally. While \texttt{pox} is in the same vein of both the \texttt{os} and \texttt{shutil} built-in modules, the majority of its functionality is unique and compliments these two modules.

\texttt{pox} provides Python equivalents of several unix shell commands such as ``which'' and ``find''. These commands allow automated discovery of what has been installed on an operating system, and where the essential tools are located. This capability is useful not only for exploring remote hosts, but also locally as a helper utility for automated build and installation.

Several high-level operations on files and filesystems are also provided. Examples of which are: finding the location of an installed Python package, determining if and where the source code resides on the filesystem, and determining what version the installed package is.

\texttt{pox} also provides utilities to enable the abstraction of commands sent to a remote filesystem. In conjunction with a registry of environment variables and installed utilites, \texttt{pox} enables the user to interact with a remote filesystem as if they were logged in locally.


\subsection*{Distributed Staging and Launching%
  \phantomsection%
  \addcontentsline{toc}{subsection}{Distributed Staging and Launching}%
  \label{distributed-staging-and-launching}%
}

\texttt{pathos} provides methods for configuring, launching, monitoring, and controlling a service on a remote host. One of the most basic features of \texttt{pathos} is the ability to configure and launch a IPC-based service on a remote host. \texttt{pathos} seeds the remote host with a small \texttt{portpicker} script, which allows the remote host to inform the localhost of a port that is available for communication.

Beyond the ability to establish a IPC service, and then post requests, is the ability to launch code in parallel. Unlike parallel computing performed at the node level (typically with MPI), \texttt{pathos} enables the user to launch jobs in parallel across heterogeneous distributed resources. \texttt{pathos} provides a distributed map-reduce algorithm, where a mix of local processors and distributed IPC services can be selected. \texttt{pathos} also provides a very basic automated load balancing service, as well as the ability for the user to directly select the resources.

A high-level interface is provided which yields a map-reduce implementation that hides the IPC internals from the user. For example, with \texttt{ipcPool}, the user can launch their code as a distributed parallel service, using standard Python and without writing a line of server or parallel batch code. \texttt{pathos} also provides tools to build a custom \texttt{Map}. In following code, the map is configured to \texttt{'autodetect'} the number of processors, and only run on the localhost:\begin{Verbatim}[commandchars=\\\{\},fontsize=\footnotesize]
\PY{c}{\PYZsh{} configure and build map}
\PY{k+kn}{from} \PY{n+nn}{pathos.launchers} \PY{k+kn}{import} \PY{n}{ipc}
\PY{k+kn}{from} \PY{n+nn}{pathos.strategies} \PY{k+kn}{import} \PY{n}{pool}
\PY{k+kn}{from} \PY{n+nn}{pathos.tools} \PY{k+kn}{import} \PY{n}{mapFactory}
\PY{n}{my\PYZus{}map} \PY{o}{=} \PY{n}{mapFactory}\PY{p}{(}\PY{n}{launcher}\PY{o}{=}\PY{n}{ipc}\PY{p}{,} \PY{n}{strategy}\PY{o}{=}\PY{n}{pool}\PY{p}{)}
\end{Verbatim}

IPC servers and communication in general is known to be insecure. However, instead of attempting to make the IPC communication itself secure, \texttt{pathos} provides the ability to automatically wrap any distributes service or communication in an SSH tunnel. SSH is a universally trusted method. Using \texttt{sshTunnel}, \texttt{pathos} has launched several distributed calculations on clusters at National Laboratories, and to date has performed test calculations that utilize node-to-node communication between two national lab clusters and a user's laptop. \texttt{pathos} allows the user to configure and launch at a very atomistic level, through raw access to \texttt{ssh} and \texttt{scp}.  Any distributed service can be tunneled, therefore less-secure methods of communication can be provided with secure authentication:\begin{Verbatim}[commandchars=\\\{\},fontsize=\footnotesize]
\PY{c}{\PYZsh{} establish a tunnel}
\PY{k+kn}{from} \PY{n+nn}{pathos.tunnel} \PY{k+kn}{import} \PY{n}{sshTunnel}
\PY{n}{uid} \PY{o}{=} \PY{l+s}{'}\PY{l+s}{foo.caltech.edu:12345:tunnel69}\PY{l+s}{'}
\PY{n}{tunnel\PYZus{}proxy} \PY{o}{=} \PY{n}{sshTunnel}\PY{p}{(}\PY{n}{uid}\PY{p}{)}

\PY{c}{\PYZsh{} inspect the ports}
\PY{n}{localport} \PY{o}{=} \PY{n}{tunnel\PYZus{}proxy}\PY{o}{.}\PY{n}{lport}
\PY{n}{remoteport} \PY{o}{=} \PY{n}{tunnel\PYZus{}proxy}\PY{o}{.}\PY{n}{rport}

\PY{c}{\PYZsh{} a user-provided model function}
\PY{k}{def} \PY{n+nf}{identify}\PY{p}{(}\PY{n}{x}\PY{p}{)}
  \PY{k}{return} \PY{n}{x}

\PY{c}{\PYZsh{} cast the model as a distributed service}
\PY{k+kn}{from} \PY{n+nn}{pathos.servers} \PY{k+kn}{import} \PY{n}{ipcServer}
\PY{n+nb}{id} \PY{o}{=} \PY{l+s}{'}\PY{l+s}{localhost:}\PY{l+s+si}{\PYZpc{}s}\PY{l+s}{:bug01}\PY{l+s}{'} \PY{o}{\PYZpc{}} \PY{n}{localport}
\PY{n}{my\PYZus{}proxy} \PY{o}{=} \PY{n}{ipcServer}\PY{p}{(}\PY{n}{identify}\PY{p}{,} \PY{n}{server}\PY{o}{=}\PY{n+nb}{id}\PY{p}{)}

\PY{c}{\PYZsh{} evaluate the model via tunneled proxy}
\PY{n}{y} \PY{o}{=} \PY{n}{my\PYZus{}proxy}\PY{p}{(}\PY{n}{x}\PY{p}{)}

\PY{c}{\PYZsh{} disconnect the tunnel}
\PY{n}{tunnel\PYZus{}proxy}\PY{o}{.}\PY{n}{disconnect}\PY{p}{(}\PY{p}{)}
\end{Verbatim}





\subsection*{Parallel Staging and Launching%
  \phantomsection%
  \addcontentsline{toc}{subsection}{Parallel Staging and Launching}%
  \label{parallel-staging-and-launching}%
}
\begin{quote}
The \texttt{pyina} package provides several basic tools to make MPI-based high-performance computing more accessable to the end user. The goal of \texttt{pyina} is to allow the user to extend their own code to MPI-based high-performance computing with minimal refactoring.
\end{quote}

The central element of \texttt{pyina} is the parallel map-reduce algorithm. \texttt{pyina} currently provides two strategies for executing the parallel-map, where a strategy is the algorithm for distributing the work list of jobs across the availble nodes. These strategies can be used ``in-the-raw'' (i.e. directly) to provide map-reduce to a user's own MPI-aware code. Further, \texttt{pyina} provides several map-reduce implementations that hide the MPI internals from the user. With these \texttt{Map} objects, the user can launch their code in parallel batch mode -{}- using standard Python and without ever having to write a line of Parallel Python or MPI code.

There are several ways that a user would typically launch their code in parallel -{}- directly with \texttt{mpirun} or \texttt{mpiexec}, or through the use of a scheduler such as torque or slurm. \texttt{pyina} encapsulates several of these launching mechanisms as \texttt{Launchers}, and provides a common interface to the different methods of launching a MPI job.  In the following code, a custom \texttt{Map} is built to execute MPI locally (i.e. not to a scheduler) using the \texttt{carddealer} strategy:\begin{Verbatim}[commandchars=\\\{\},fontsize=\footnotesize]
\PY{c}{\PYZsh{} configure and build map}
\PY{k+kn}{from} \PY{n+nn}{pyina.launchers} \PY{k+kn}{import} \PY{n}{mpirun}
\PY{k+kn}{from} \PY{n+nn}{pyina.strategies} \PY{k+kn}{import} \PY{n}{carddealer} \PY{k}{as} \PY{n}{card}
\PY{k+kn}{from} \PY{n+nn}{pyina.tools} \PY{k+kn}{import} \PY{n}{mapFactory}
\PY{n}{my\PYZus{}map} \PY{o}{=} \PY{n}{mapFactory}\PY{p}{(}\PY{l+m+mi}{4}\PY{p}{,} \PY{n}{launcher}\PY{o}{=}\PY{n}{mpirun}\PY{p}{,} \PY{n}{strategy}\PY{o}{=}\PY{n}{card}\PY{p}{)}
\end{Verbatim}




\subsection*{New Massively-Parallel Optimization Algorithms%
  \phantomsection%
  \addcontentsline{toc}{subsection}{New Massively-Parallel Optimization Algorithms}%
  \label{new-massively-parallel-optimization-algorithms}%
}
In \texttt{mystic}, optimizers have been extended to parallel whenever possible. To have an optimizer execute in parallel, the user only needs to provide the solver with a parallel map.  For example, extending the Differential Evolution {[}\hyperref[skp95]{SKP95}{]} solver to parallel is involves passing a \texttt{Map} to the \texttt{SetEvaluationMap} method. In the example below, each generation has $20$ candidates, and will execute in parallel using MPI with $4$ workers:\begin{Verbatim}[commandchars=\\\{\},fontsize=\footnotesize]
\PY{c}{\PYZsh{} the function to be minimized and the bounds}
\PY{k+kn}{from} \PY{n+nn}{mystic.models} \PY{k+kn}{import} \PY{n}{rosen} \PY{k}{as} \PY{n}{my\PYZus{}model}
\PY{n}{lb} \PY{o}{=} \PY{p}{[}\PY{l+m+mf}{0.0}\PY{p}{,} \PY{l+m+mf}{0.0}\PY{p}{,} \PY{l+m+mf}{0.0}\PY{p}{]}
\PY{n}{ub} \PY{o}{=} \PY{p}{[}\PY{l+m+mf}{2.0}\PY{p}{,} \PY{l+m+mf}{2.0}\PY{p}{,} \PY{l+m+mf}{2.0}\PY{p}{]}

\PY{c}{\PYZsh{} get termination condition object}
\PY{k+kn}{from} \PY{n+nn}{mystic.termination} \PY{k+kn}{import} \PY{n}{ChangeOverGeneration}
\PY{n}{COG} \PY{o}{=} \PY{n}{ChangeOverGeneration}\PY{p}{(}\PY{p}{)}

\PY{c}{\PYZsh{} select the parallel launch configuration}
\PY{k+kn}{from} \PY{n+nn}{pyina.maps} \PY{k+kn}{import} \PY{n}{MpirunCarddealer}
\PY{n}{my\PYZus{}map} \PY{o}{=} \PY{n}{MpirunCarddealer}\PY{p}{(}\PY{l+m+mi}{4}\PY{p}{)}

\PY{c}{\PYZsh{} instantiate and configure the solver}
\PY{k+kn}{from} \PY{n+nn}{mystic.solvers} \PY{k+kn}{import} \PY{n}{DifferentialEvolutionSolver}
\PY{n}{solver} \PY{o}{=} \PY{n}{DifferentialEvolutionSolver}\PY{p}{(}\PY{n+nb}{len}\PY{p}{(}\PY{n}{lb}\PY{p}{)}\PY{p}{,} \PY{l+m+mi}{20}\PY{p}{)}
\PY{n}{solver}\PY{o}{.}\PY{n}{SetRandomInitialPoints}\PY{p}{(}\PY{n}{lb}\PY{p}{,} \PY{n}{ub}\PY{p}{)}
\PY{n}{solver}\PY{o}{.}\PY{n}{SetStrictRanges}\PY{p}{(}\PY{n}{lb}\PY{p}{,} \PY{n}{ub}\PY{p}{)}
\PY{n}{solver}\PY{o}{.}\PY{n}{SetEvaluationMap}\PY{p}{(}\PY{n}{my\PYZus{}map}\PY{p}{)}
\PY{n}{solver}\PY{o}{.}\PY{n}{Solve}\PY{p}{(}\PY{n}{my\PYZus{}model}\PY{p}{,} \PY{n}{COG}\PY{p}{)}

\PY{c}{\PYZsh{} obtain the solution}
\PY{n}{solution} \PY{o}{=} \PY{n}{solver}\PY{o}{.}\PY{n}{bestSolution}
\end{Verbatim}
\begin{figure}[]
\noindent{\includegraphics[width=\columnwidth]{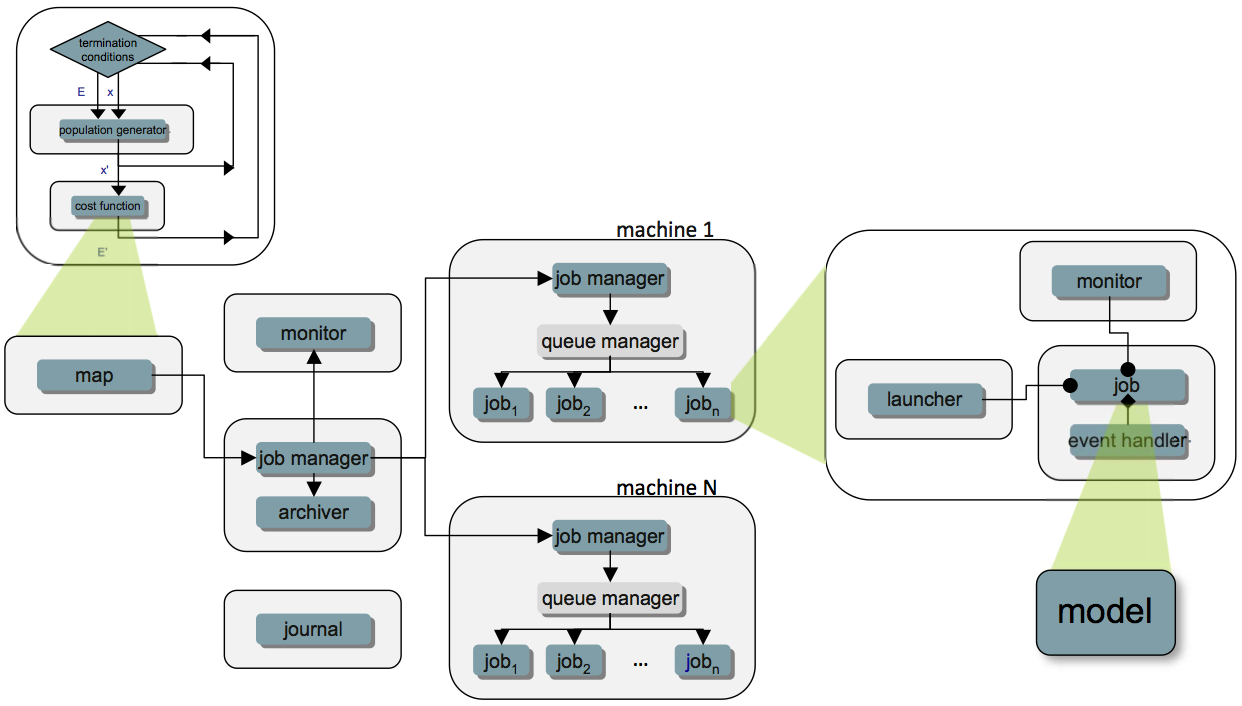}\hfill}
\caption{Conceptual diagram for a \texttt{carddealer-DE} optimizer. The optimizer contains a map function that stages $n$ copies of the user's model $F(x)$ in parallel across distributed resources. \DUrole{label}{fig-desolver}}
\end{figure}
Another type of new parallel solver utilizes the \texttt{SetNestedSolver} method to stage a parallel launch of $N$ optimizers, each with different initial conditions. The following code shows the \texttt{BuckshotSolver} scheduling a launch of $N=20$ optimizers in parallel to the default queue, where $5$ nodes each with $4$ processors have been requested:\begin{Verbatim}[commandchars=\\\{\},fontsize=\footnotesize]
\PY{c}{\PYZsh{} the function to be minimized and the bounds}
\PY{k+kn}{from} \PY{n+nn}{mystic.models} \PY{k+kn}{import} \PY{n}{rosen} \PY{k}{as} \PY{n}{my\PYZus{}model}
\PY{n}{lb} \PY{o}{=} \PY{p}{[}\PY{l+m+mf}{0.0}\PY{p}{,} \PY{l+m+mf}{0.0}\PY{p}{,} \PY{l+m+mf}{0.0}\PY{p}{]}
\PY{n}{ub} \PY{o}{=} \PY{p}{[}\PY{l+m+mf}{2.0}\PY{p}{,} \PY{l+m+mf}{2.0}\PY{p}{,} \PY{l+m+mf}{2.0}\PY{p}{]}

\PY{c}{\PYZsh{} get monitor and termination condition objects}
\PY{k+kn}{from} \PY{n+nn}{mystic.monitors} \PY{k+kn}{import} \PY{n}{LoggingMonitor}
\PY{n}{stepmon} \PY{o}{=} \PY{n}{LoggingMonitor}\PY{p}{(}\PY{l+m+mi}{1}\PY{p}{,} \PY{l+s}{'}\PY{l+s}{log.txt}\PY{l+s}{'}\PY{p}{)}
\PY{k+kn}{from} \PY{n+nn}{mystic.termination} \PY{k+kn}{import} \PY{n}{ChangeOverGeneration}
\PY{n}{COG} \PY{o}{=} \PY{n}{ChangeOverGeneration}\PY{p}{(}\PY{p}{)}

\PY{c}{\PYZsh{} select the parallel launch configuration}
\PY{k+kn}{from} \PY{n+nn}{pyina.maps} \PY{k+kn}{import} \PY{n}{TorqueMpirunCarddealer}
\PY{n}{my\PYZus{}map} \PY{o}{=} \PY{n}{TorqueMpirunCarddealer}\PY{p}{(}\PY{l+s}{'}\PY{l+s}{5:ppn=4}\PY{l+s}{'}\PY{p}{)}

\PY{c}{\PYZsh{} instantiate and configure the nested solver}
\PY{k+kn}{from} \PY{n+nn}{mystic.solvers} \PY{k+kn}{import} \PY{n}{PowellDirectionalSolver}
\PY{n}{my\PYZus{}solver} \PY{o}{=} \PY{n}{PowellDirectionalSolver}\PY{p}{(}\PY{n+nb}{len}\PY{p}{(}\PY{n}{lb}\PY{p}{)}\PY{p}{)}
\PY{n}{my\PYZus{}solver}\PY{o}{.}\PY{n}{SetStrictRanges}\PY{p}{(}\PY{n}{lb}\PY{p}{,} \PY{n}{ub}\PY{p}{)}
\PY{n}{my\PYZus{}solver}\PY{o}{.}\PY{n}{SetEvaluationLimits}\PY{p}{(}\PY{l+m+mi}{50}\PY{p}{)}

\PY{c}{\PYZsh{} instantiate and configure the outer solver}
\PY{k+kn}{from} \PY{n+nn}{mystic.solvers} \PY{k+kn}{import} \PY{n}{BuckshotSolver}
\PY{n}{solver} \PY{o}{=} \PY{n}{BuckshotSolver}\PY{p}{(}\PY{n+nb}{len}\PY{p}{(}\PY{n}{lb}\PY{p}{)}\PY{p}{,} \PY{l+m+mi}{20}\PY{p}{)}
\PY{n}{solver}\PY{o}{.}\PY{n}{SetRandomInitialPoints}\PY{p}{(}\PY{n}{lb}\PY{p}{,} \PY{n}{ub}\PY{p}{)}
\PY{n}{solver}\PY{o}{.}\PY{n}{SetGenerationMonitor}\PY{p}{(}\PY{n}{stepmon}\PY{p}{)}
\PY{n}{solver}\PY{o}{.}\PY{n}{SetNestedSolver}\PY{p}{(}\PY{n}{my\PYZus{}solver}\PY{p}{)}
\PY{n}{solver}\PY{o}{.}\PY{n}{SetSolverMap}\PY{p}{(}\PY{n}{my\PYZus{}map}\PY{p}{)}
\PY{n}{solver}\PY{o}{.}\PY{n}{Solve}\PY{p}{(}\PY{n}{my\PYZus{}model}\PY{p}{,} \PY{n}{COG}\PY{p}{)}

\PY{c}{\PYZsh{} obtain the solution}
\PY{n}{solution} \PY{o}{=} \PY{n}{solver}\PY{o}{.}\PY{n}{bestSolution}
\end{Verbatim}



\begin{figure}[]
\noindent{\includegraphics[width=\columnwidth]{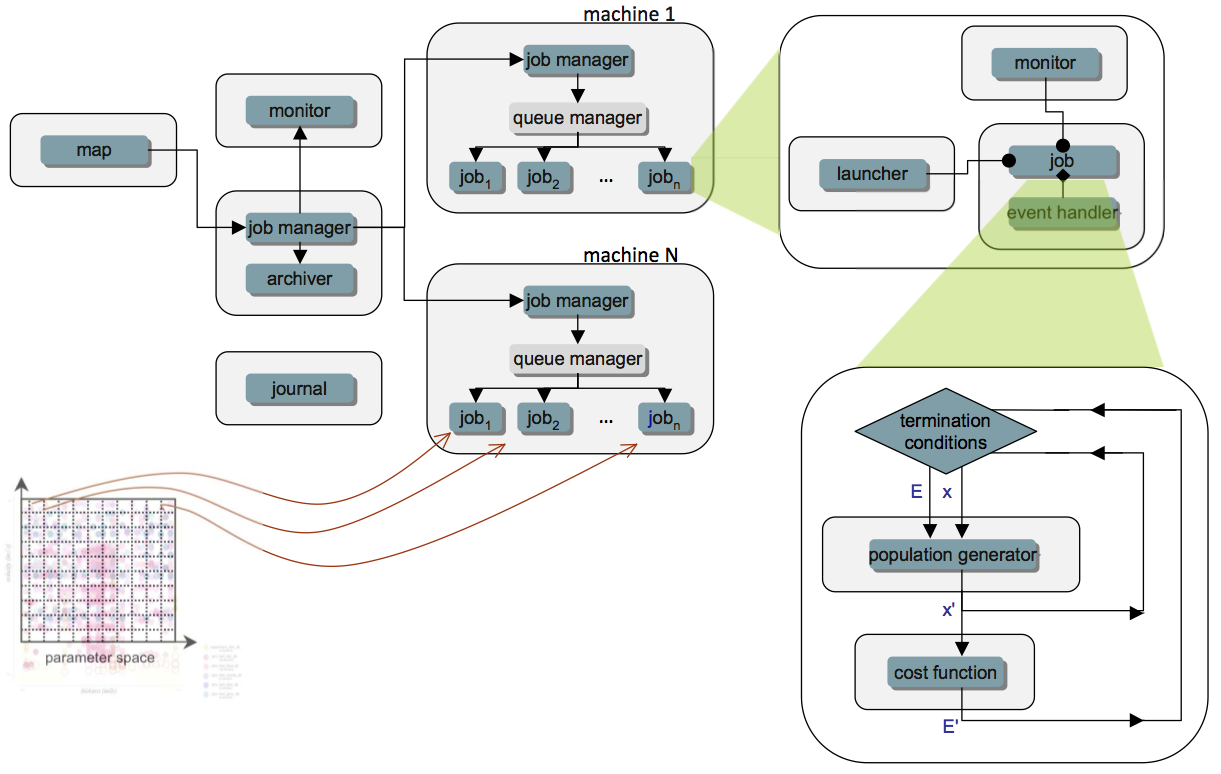}\hfill}
\caption{Conceptual diagram for a \texttt{lattice-Powell} optimizer. $N$ Powell's local-search optimizers are launched in parallel, with each optimizer starting from the center of a different lattice cuboid in parameter space. A \texttt{buckshot-Powell} optimizer is similar; however, instead utilizes a uniform random distribution of initial values.  \DUrole{label}{fig-batchgrid}}
\end{figure}




\subsection*{Probability and Uncertainty Tooklit%
  \phantomsection%
  \addcontentsline{toc}{subsection}{Probability and Uncertainty Tooklit}%
  \label{probability-and-uncertainty-tooklit}%
}
The software framework presented in this paper was designed to solve UQ problems. Calculation of the upper and lower bounds for probability of failure is provided as a framework service. The McDiarmid subdiameter is a model-based measure of sensitivity, and is cast within \texttt{mystic} as a global optimization. Diameter calculations can be coupled with partitioning algorithms, and used to discover regions of critical behavior. Optimization over probability measures is also available as a framework service, and is utilized in (OUQ) calculations of optimal bounds.

The minimization or maximization of a cost function is the basis for performing most calculations in \texttt{mystic}. The optimizer generates new trial parameters, which are evaluated in a user-provided model function against a user-provided metric. Two simple difference metrics provided are:  $metric = | F(x) - G |^2$, where $F$ is the model function evaluated at some trial set of fit parameters $\cal P$, and $G$ is the corresponding experimental data -{}- and $metric = | F(x) - F(y) |^2$, where $x$ and $y$ are two slightly different sets of input parameters (\DUrole{ref}{eqn-subosc}).

\texttt{mystic} provides factory methods to automate the generation of a cost function from a user's model. Conceptually, a \texttt{costFactory} is as follows:\begin{Verbatim}[commandchars=\\\{\},fontsize=\footnotesize]
\PY{c}{\PYZsh{} prepare a (F(X) - G)**2 a metric}
\PY{k}{def} \PY{n+nf}{costFactory}\PY{p}{(}\PY{n}{my\PYZus{}model}\PY{p}{,} \PY{n}{my\PYZus{}data}\PY{p}{)}\PY{p}{:}
  \PY{k}{def} \PY{n+nf}{cost}\PY{p}{(}\PY{n}{param}\PY{p}{)}\PY{p}{:}

    \PY{c}{\PYZsh{} compute the cost}
    \PY{k}{return} \PY{p}{(} \PY{n}{my\PYZus{}model}\PY{p}{(}\PY{n}{param}\PY{p}{)} \PY{o}{-} \PY{n}{my\PYZus{}data} \PY{p}{)}\PY{o}{*}\PY{o}{*}\PY{l+m+mi}{2}

  \PY{k}{return} \PY{n}{cost}
\end{Verbatim}

\begin{figure}[]
\noindent{\includegraphics[width=\columnwidth]{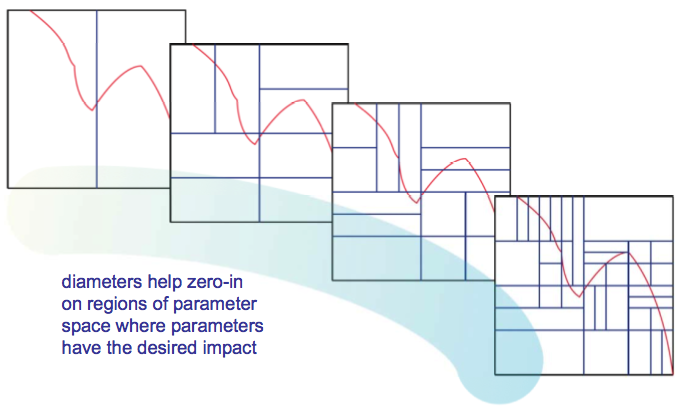}\hfill}
\caption{Coupling an iterative partitioning algorithm with a sensitivity calculation enables the discovery of critical regions in parameter space. \DUrole{label}{fig-partitioning}}
\end{figure}
Suboscillations (\DUrole{ref}{eqn-subosc}), used in calculations of rigorous sensitivity (such as $D_{i} / D$), can also be cast as a cost metric:\begin{Verbatim}[commandchars=\\\{\},fontsize=\footnotesize]
\PY{c}{\PYZsh{} prepare a (F(X) - F(X'))**2 cost metric}
\PY{k}{def} \PY{n+nf}{suboscillationFactory}\PY{p}{(}\PY{n}{my\PYZus{}model}\PY{p}{,} \PY{n}{i}\PY{p}{)}\PY{p}{:}

  \PY{k}{def} \PY{n+nf}{cost}\PY{p}{(}\PY{n}{param}\PY{p}{)}\PY{p}{:}

    \PY{c}{\PYZsh{} get X and X' (Xi' is appended to X at param[-1])}
    \PY{n}{x}       \PY{o}{=} \PY{n}{param}\PY{p}{[}\PY{p}{:}\PY{o}{-}\PY{l+m+mi}{1}\PY{p}{]}
    \PY{n}{x\PYZus{}prime} \PY{o}{=} \PY{n}{param}\PY{p}{[}\PY{p}{:}\PY{n}{i}\PY{p}{]} \PY{o}{+} \PY{n}{param}\PY{p}{[}\PY{o}{-}\PY{l+m+mi}{1}\PY{p}{:}\PY{p}{]} \PY{o}{+} \PY{n}{param}\PY{p}{[}\PY{n}{i}\PY{o}{+}\PY{l+m+mi}{1}\PY{p}{:}\PY{o}{-}\PY{l+m+mi}{1}\PY{p}{]}

    \PY{c}{\PYZsh{} compute the suboscillation}
    \PY{k}{return} \PY{o}{-}\PY{p}{(} \PY{n}{my\PYZus{}model}\PY{p}{(}\PY{n}{x}\PY{p}{)} \PY{o}{-} \PY{n}{my\PYZus{}model}\PY{p}{(}\PY{n}{x\PYZus{}prime}\PY{p}{)} \PY{p}{)}\PY{o}{*}\PY{o}{*}\PY{l+m+mi}{2}

  \PY{k}{return} \PY{n}{cost}
\end{Verbatim}
The diameter $D$ (\DUrole{ref}{eqn-diamosc}) is the root-mean square of its component-wise suboscillations.  The calculation of the diameter is performed as a nested optimization, as shown above for the \texttt{BuckshotSolver}. Each inner optimization is a calculation of a component suboscillation, using the a global optimizer (such as \texttt{DifferentialEvolutionSolver}) and the cost metric shown above.

The optimization algorithm takes a set of model parameters $\cal P$ and the current measure of oscillation $O({\cal P})$ as inputs, and produces an updated $\cal P$. The optimization loop iterates until the termination conditions are satisfied.

When the global optimization terminates the condition $O({\cal P}) < -(osc^2_i + \epsilon)$ is satisfied, and the final set ${\cal P}$ is composed of $X$ and ${X}'$.


OUQ problems can be thought of optimization problems where the goal is to find the global maximum of a probability function $\mu[H \leq 0]$, where $H \leq 0$ is a failure criterion for the model response function $H$.  Additional conditions in an OUQ problem are provided as constraints on the information set. Typically, a condition such as a mean constraint on $H$, $m_{1} \leq \mathbb{E}_{\mu}[H] \leq m_{2}$, will be imposed on the maximization.  After casting the OUQ problem in terms of optimization and constraints, we can plug these terms into the infrastructure provided by \texttt{mystic}.

Optimal uncertainty quantification (OUQ) is maximization over a probability distribution, and not over a standard difference metric. Therefore the fundamental data structure is not the user-provided model function, but is a user-configured probability measure. For example, a discrete measure is represented by a collection of support points, each with an accompanying weight. Measures come with built-in methods for calculating the mass, range, and mean of the measure, and also for imposing a mass, range, and mean on the measure. Measures also have some very basic operations, including point addition and subtraction, and the formation of product measures.
\begin{figure}[]
\noindent{\includegraphics[width=\columnwidth]{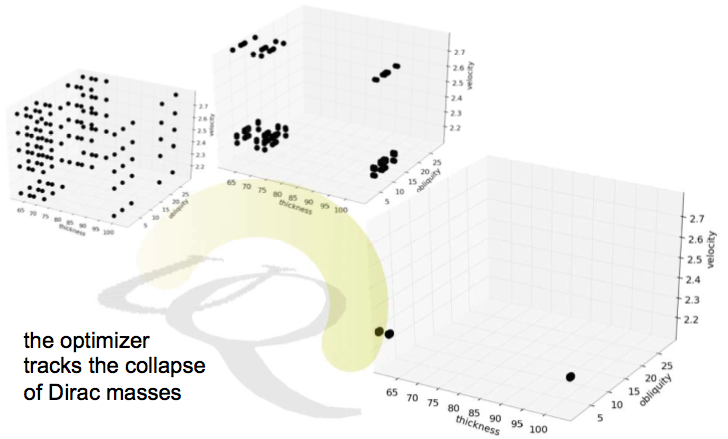}\hfill}
\caption{Optimal uncertainty quantification is an optimization of probability measures over design parameter space. Collapse of probability masses corresponds to the determination of the critical design parameters. \DUrole{label}{fig-ouq}}
\end{figure}

Global optimizations used in solving OUQ problems are composed in the same manner as shown above for the \texttt{DifferentialEvolutionSolver}. The cost function, however, is not formulated as in the examples above -{}- OUQ is an optimization over product measures, and thus uses \texttt{mystic}'s \texttt{product\_measure} class as the target of the optimization.  Also as shown above, the bounds constraints are imposed with the \texttt{SetStrictRanges} method, while parameter constraints (composed as below) are imposed with the \texttt{SetConstraints} method. The union set of these constraints defines the set $\mathcal{A}$.


So for example, let us define the feasable set\begin{equation}
\label{eqn-mathcala}
\mathcal{A} = \left\{ (f, \mu) \,\middle|\,
    \begin{matrix}
        f = \tt{my\_model} \,:\, \prod_{i=1}^{3} [\tt{lb}_{i}, \tt{ub}_{i}] \to \mathbb{R}, \\
        \mu = \bigotimes_{i=1}^{3} \mu_{i} \in \bigotimes_{i=1}^{3} \mathcal{M}([\tt{lb}_{i}, \tt{ub}_{i}]), \\
        \tt{m}_{\tt{lb}} \leq \mathbb{E}_{\mu}[f] \leq \tt{m}_{\tt{ub}}
    \end{matrix} \right\}
\end{equation}which reduces to the finite-dimensional subset\begin{equation}
\label{eqn-mathcaladelta}
\mathcal{A}_{\Delta} = \left\{ (f, \mu) \in \mathcal{A} \,\middle|\,
    \begin{matrix}
        \text{for } \vec{x} \text{ and } \vec{y} \in \prod_{i=1}^{3} [\tt{lb}_{i}, \tt{ub}_{i}], \\
        \text{and } \vec{w} \in [0, 1], \\
        \mu_{i} = w_{i} \delta_{x_{i}} + (1 - w_{i}) \delta_{y_{i}}
    \end{matrix} \right\}
\end{equation}where $\vec{x} = \text{some }(x_{1}, x_{2}, x_{3})$,
$\vec{y} = \text{some }(y_{1}, y_{2}, y_{3})$,
and $\vec{w} = \text{some }(w_{1}, w_{2}, w_{3})$.

To solve this OUQ problem, we first write the code
for the bounds, cost function, and constraints -{}- then
we plug this code into a global optimization script,
as noted above.

OUQ requires the user provide a list of bounds that follow the
formatting convention that \texttt{mystic}'s \texttt{product\_measure.load}
uses to build a product measure from a list of input parameters.
This roughly follows the definition of a product measure as
shown in equation (\DUrole{ref}{eqn-mathcaladelta}),
and also is detailed in the comment block below:\begin{Verbatim}[commandchars=\\\{\},fontsize=\footnotesize]
\PY{c}{\PYZsh{} OUQ requires bounds in a very specific form...}
\PY{c}{\PYZsh{} param = [wxi]*nx + [xi]*nx \PYZbs{}}
\PY{c}{\PYZsh{}       + [wyi]*ny + [yi]*ny \PYZbs{}}
\PY{c}{\PYZsh{}       + [wzi]*nz + [zi]*nz}
\PY{n}{npts} \PY{o}{=} \PY{p}{(}\PY{n}{nx}\PY{p}{,}\PY{n}{ny}\PY{p}{,}\PY{n}{nz}\PY{p}{)}
\PY{n}{lb} \PY{o}{=} \PY{p}{(}\PY{n}{nx} \PY{o}{*} \PY{n}{w\PYZus{}lower}\PY{p}{)} \PY{o}{+} \PY{p}{(}\PY{n}{nx} \PY{o}{*} \PY{n}{x\PYZus{}lower}\PY{p}{)} \PYZbs{}
   \PY{o}{+} \PY{p}{(}\PY{n}{ny} \PY{o}{*} \PY{n}{w\PYZus{}lower}\PY{p}{)} \PY{o}{+} \PY{p}{(}\PY{n}{ny} \PY{o}{*} \PY{n}{y\PYZus{}lower}\PY{p}{)} \PYZbs{}
   \PY{o}{+} \PY{p}{(}\PY{n}{nz} \PY{o}{*} \PY{n}{w\PYZus{}lower}\PY{p}{)} \PY{o}{+} \PY{p}{(}\PY{n}{nz} \PY{o}{*} \PY{n}{z\PYZus{}lower}\PY{p}{)}
\PY{n}{ub} \PY{o}{=} \PY{p}{(}\PY{n}{nx} \PY{o}{*} \PY{n}{w\PYZus{}upper}\PY{p}{)} \PY{o}{+} \PY{p}{(}\PY{n}{nx} \PY{o}{*} \PY{n}{x\PYZus{}upper}\PY{p}{)} \PYZbs{}
   \PY{o}{+} \PY{p}{(}\PY{n}{ny} \PY{o}{*} \PY{n}{w\PYZus{}upper}\PY{p}{)} \PY{o}{+} \PY{p}{(}\PY{n}{ny} \PY{o}{*} \PY{n}{y\PYZus{}upper}\PY{p}{)} \PYZbs{}
   \PY{o}{+} \PY{p}{(}\PY{n}{nz} \PY{o}{*} \PY{n}{w\PYZus{}upper}\PY{p}{)} \PY{o}{+} \PY{p}{(}\PY{n}{nz} \PY{o}{*} \PY{n}{z\PYZus{}upper}\PY{p}{)}
\end{Verbatim}
The constraints function and the cost function
typically require the use of measure mathematics.
In the example below, the constraints check if
\texttt{measure.mass} $\approx 1.0$; if not,
the the measure's mass is normalized to $1.0$.
The second block of constraints below check if
$m_{1} \leq \mathbb{E}_{\mu}[H] \leq m_{2}$,
where $m_{1} =$ \texttt{target\_mean} $-$ \texttt{error}
and $m_{2} =$ \texttt{target\_mean} $+$ \texttt{error};
if not, an optimization is performed to satisfy
this mean constraint.
The \texttt{product\_measure} is built (with \texttt{load})
from the optimization parameters \texttt{param}, and
after all the constraints are applied, \texttt{flatten}
is used to extract the updated \texttt{param}:\begin{Verbatim}[commandchars=\\\{\},fontsize=\footnotesize]
\PY{k+kn}{from} \PY{n+nn}{mystic.math.measures} \PY{k+kn}{import} \PY{n}{split\PYZus{}param}
\PY{k+kn}{from} \PY{n+nn}{mystic.math.dirac\PYZus{}measure} \PY{k+kn}{import} \PY{n}{product\PYZus{}measure}
\PY{k+kn}{from} \PY{n+nn}{mystic.math} \PY{k+kn}{import} \PY{n}{almostEqual}

\PY{c}{\PYZsh{} split bounds into weight-only & sample-only}
\PY{n}{w\PYZus{}lb}\PY{p}{,} \PY{n}{m\PYZus{}lb} \PY{o}{=} \PY{n}{split\PYZus{}param}\PY{p}{(}\PY{n}{lb}\PY{p}{,} \PY{n}{npts}\PY{p}{)}
\PY{n}{w\PYZus{}ub}\PY{p}{,} \PY{n}{m\PYZus{}ub} \PY{o}{=} \PY{n}{split\PYZus{}param}\PY{p}{(}\PY{n}{ub}\PY{p}{,} \PY{n}{npts}\PY{p}{)}

\PY{c}{\PYZsh{} generate constraints function}
\PY{k}{def} \PY{n+nf}{constraints}\PY{p}{(}\PY{n}{param}\PY{p}{)}\PY{p}{:}
  \PY{n}{prodmeasure} \PY{o}{=} \PY{n}{product\PYZus{}measure}\PY{p}{(}\PY{p}{)}
  \PY{n}{prodmeasure}\PY{o}{.}\PY{n}{load}\PY{p}{(}\PY{n}{param}\PY{p}{,} \PY{n}{npts}\PY{p}{)}

  \PY{c}{\PYZsh{} impose norm on measures}
  \PY{k}{for} \PY{n}{measure} \PY{o+ow}{in} \PY{n}{prodmeasure}\PY{p}{:}
    \PY{k}{if} \PY{o+ow}{not} \PY{n}{almostEqual}\PY{p}{(}\PY{n+nb}{float}\PY{p}{(}\PY{n}{measure}\PY{o}{.}\PY{n}{mass}\PY{p}{)}\PY{p}{,} \PY{l+m+mf}{1.0}\PY{p}{)}\PY{p}{:}
      \PY{n}{measure}\PY{o}{.}\PY{n}{normalize}\PY{p}{(}\PY{p}{)}

  \PY{c}{\PYZsh{} impose expectation on product measure}
  \PY{n}{E} \PY{o}{=} \PY{n+nb}{float}\PY{p}{(}\PY{n}{prodmeasure}\PY{o}{.}\PY{n}{get\PYZus{}expect}\PY{p}{(}\PY{n}{my\PYZus{}model}\PY{p}{)}\PY{p}{)}
  \PY{k}{if} \PY{o+ow}{not} \PY{p}{(}\PY{n}{E} \PY{o}{<}\PY{o}{=} \PY{n+nb}{float}\PY{p}{(}\PY{n}{target\PYZus{}mean} \PY{o}{+} \PY{n}{error}\PY{p}{)}\PY{p}{)} \PYZbs{}
  \PY{o+ow}{or} \PY{o+ow}{not} \PY{p}{(}\PY{n+nb}{float}\PY{p}{(}\PY{n}{target\PYZus{}mean} \PY{o}{-} \PY{n}{error}\PY{p}{)} \PY{o}{<}\PY{o}{=} \PY{n}{E}\PY{p}{)}\PY{p}{:}
    \PY{n}{prodmeasure}\PY{o}{.}\PY{n}{set\PYZus{}expect}\PY{p}{(}\PY{p}{(}\PY{n}{target\PYZus{}mean}\PY{p}{,} \PY{n}{error}\PY{p}{)}\PY{p}{,} \PYZbs{}
                           \PY{n}{my\PYZus{}model}\PY{p}{,} \PY{p}{(}\PY{n}{m\PYZus{}lb}\PY{p}{,} \PY{n}{m\PYZus{}ub}\PY{p}{)}\PY{p}{)}

  \PY{c}{\PYZsh{} extract weights and positions}
  \PY{k}{return} \PY{n}{prodmeasure}\PY{o}{.}\PY{n}{flatten}\PY{p}{(}\PY{p}{)}
\end{Verbatim}
The PoF is calculated in the cost function with the \texttt{pof} method:\begin{Verbatim}[commandchars=\\\{\},fontsize=\footnotesize]
\PY{c}{\PYZsh{} generate maximizing function}
\PY{k}{def} \PY{n+nf}{cost}\PY{p}{(}\PY{n}{param}\PY{p}{)}\PY{p}{:}
  \PY{n}{prodmeasure} \PY{o}{=} \PY{n}{product\PYZus{}measure}\PY{p}{(}\PY{p}{)}
  \PY{n}{prodmeasure}\PY{o}{.}\PY{n}{load}\PY{p}{(}\PY{n}{param}\PY{p}{,} \PY{n}{npts}\PY{p}{)}
  \PY{k}{return} \PY{n}{MINMAX} \PY{o}{*} \PY{n}{prodmeasure}\PY{o}{.}\PY{n}{pof}\PY{p}{(}\PY{n}{my\PYZus{}model}\PY{p}{)}
\end{Verbatim}
We find the \emph{supremum} (as in \DUrole{ref}{eqn-ouqgeneral}) when \texttt{MINMAX=-1} and, upon solution, the function maximum is \texttt{-solver.bestEnergy}. We find the \emph{infimum} when \texttt{MINMAX=1} and, upon solution, the function minimum is \texttt{solver.bestEnergy}.


\subsection*{Future Developments%
  \phantomsection%
  \addcontentsline{toc}{subsection}{Future Developments}%
  \label{future-developments}%
}

Many of the features presented above are not currently in released versions of the code. Of primary importance is to migrate these features from development branches to a new release.

The next natural question beyond ``what is the sensitivity of a model to an input parameter?'' is ``how does the correlation between input parameters affect the outcome of the model?''. Methods for calculating parameter correlation will be very useful in analysis of results. Another natural question is how to handle uncertainty in the data.

New partitioning algorithms for the discovery of regions of critical behavior will be added to \texttt{mystic}. Currently the only partitioning rule drives the optimizer toward partitioning space such that the upper bounds of a ``piecewise-McDiarmid'' type are iteratively tightened {[}\hyperref[stm11]{STM11}{]}. We will extend the partitioning algorithm not to refine the diameter, but to discover regions where the diameters meet a set of criteria (such as: regions where there are two roughly equal subdiameters that account for 90\% or more of the total diameter (i.e. automated discovery of regions where two parameters compete to govern the system behavior). \texttt{mystic} will also further expand its base of available statistical and measure methods, equation solvers, and also make available several more traditional uncertainty quantification methods. \texttt{mystic} will continue to expand its base of optimizers, with particular emphasis on new optimization algorithms that efficiently utilize parallel computing. \texttt{mystic} currently has a few simple parallel optimization algorithms, such as the \texttt{LatticeSolver} and \texttt{BuckshotSolver} solvers; however, algorithms that utilize a variant of game theory to do speculation about future iterations (i.e. break the paradigm of an iteration being a blocker to parallelism), or use parallelism and dynamic constraints to allow optimizers launched in parallel to avoid finding the same minimum twice, are planned. Parallelism in optimization also allows us to not only find the global minima, but to simultaneously find all local minima and transition points -{}- thus providing a much more efficient means of mapping out a potential energy surface. Solving uncertainty quantification problems requires a lot of computational resources and often must require a minimum of both model evaluations and accompanying experiments, so we also have to keep an eye on developing parallel algorithms for global optimization with overall computational efficiency.

\texttt{pathos} includes utilities for filesystem exploration and automated builds, and a utility for the serialization of Python objects, however these framework services will need to be made more robust as more platforms and more extensive objects and codes are tackled. Effort will continue on expanding the management and platform capabilities for \texttt{pathos}, unifying and hardening the map interface and providing load balancing for all types of connections. The high-level interface to analysis circuits will be extended to encompass new types of logic for combining and nesting components (as nested optimizers are utilized in many materials theory codes). Monitoring and logging to files and databases across parallel and distributed resources will be migrated from \texttt{mystic} and added as \texttt{pathos} framework services.


\subsection*{Summary%
  \phantomsection%
  \addcontentsline{toc}{subsection}{Summary}%
  \label{summary}%
}

A brief overview of the mathematical and software components used in building a software framework for predictive science is presented.


\subsection*{Acknowledgements%
  \phantomsection%
  \addcontentsline{toc}{subsection}{Acknowledgements}%
  \label{acknowledgements}%
}

This material is based upon work supported by the Department of Energy National Nuclear Security Administration under Award Number DE-FC52-08NA28613, and by the National Science Foundation under Award Number DMR-0520547.

\end{document}